\documentclass{aa}

\newcommand{\Nstar}{TOI-858\,B}
\newcommand{\NstarBTIC}{TIC-198008002}
\newcommand{\Nplanet}{TOI-858\,B\,b}
\newcommand{\NstarB}{TOI-858\,A}
\providecommand{\feh}{$\left[{\rm Fe / H}\right]$}
\newcommand{\kms}{km\,s$^{-1}$}

\newcommand{\ms}{m\,s$^{-1}$}

\newcommand{\masy}{mas\,yr$^{-1}$}

\newcommand{\rpl}{R$_{p}$}
\newcommand{\teff}{$T_{\rm eff}$}

\providecommand{\bjdtdb}{\ensuremath{\rm {BJD_{TDB}}}}

\providecommand{\msun}{\ensuremath{\mathrm M_{\sun}}}
\providecommand{\rsun}{\ensuremath{\mathrm R_{\sun}}}
\providecommand{\lsun}{\ensuremath{\mathrm L_{\sun}}}
\providecommand{\mj}{\ensuremath{\mathrm M_{\rm J}}}
\providecommand{\rj}{\ensuremath{\mathrm R_{\rm J}}}

\providecommand{\fave}{\langle F \rangle}
\providecommand{\fluxcgs}{10$^9$ erg s$^{-1}$ cm$^{-2}$}

\newcommand{\LSO}{La Silla Observatory}

\newcommand{\tess}{{\small \it TESS}}

\newcommand{\gaia}{{\it Gaia}}

\newcommand{\exofast}{{\it EXOFASTv2}}
\newcommand{\emp}{\textsc{SpecMatch-emp}}

\providecommand{\mplanet}{\ensuremath{1.10 ^{+0.08}_{-0.07}\,\mj}}

\usepackage[svgnames]{xcolor}
\usepackage{placeins}
\usepackage{graphicx}
\usepackage[varg]{txfonts}
\usepackage[colorlinks,
linkcolor={blue},
citecolor={blue},
urlcolor={blue}]{hyperref}

\usepackage{orcidlink}

\bibpunct{(}{)}{;}{a}{}{,} 

\begin{document}

   \title{\Nplanet: A hot Jupiter on a polar orbit in a loose binary
 \thanks{Based on observations collected at the European Southern
        Observatory, Chile with the CORALIE echelle spectrograph mounted on the
         1.2 m Swiss  telescope  at La Silla Observatory and with ESO HARPS Open 
         Time (123.C-0123, 123.C-0123,).  }}
        
   \author{J. Hagelberg\orcidlink{0000-0002-1096-1433} 
          \inst{\ref{inst:unige}}
          \and
          L.~D. Nielsen\orcidlink{0000-0002-5254-2499}\inst{\ref{inst:unige}, \ref{inst:eso-garching}}
          \and
          O. Attia\orcidlink{0000-0002-7971-7439}\inst{\ref{inst:unige}}
          \and
          V. Bourrier\inst{\ref{inst:unige}}
          \and
          L. Pearce\inst{\ref{inst:arizona}}
          \and 
        J. Venturini\orcidlink{0000-0001-9527-2903}\inst{\ref{inst:unige}}
        \and
                  J.~N.~ Winn\orcidlink{0000-0002-4265-047X}\inst{\ref{inst:princeton}}
          \and 
          F. Bouchy\orcidlink{0000-0002-7613-393X}\inst{\ref{inst:unige}}
          \and 
          L.~G.~Bouma\orcidlink{0000-0002-0514-5538}\inst{\ref{inst:caltech}}
          \and 
          C.~Brice\~{n}o\inst{\ref{inst:ctio}}
          \and 
          K.~A.~Collins\orcidlink{0000-0001-6588-9574}\inst{\ref{inst:cfa}}
          \and 
          A.~B.~Davis\orcidlink{0000-0002-5070-8395}\inst{\ref{inst:yale}}
          \and 
          J.~D.~Eastman\orcidlink{0000-0003-3773-5142]}\inst{\ref{inst:cfa}}
          \and 
          P. Evans\orcidlink{0000-0002-5674-2404}\inst{\ref{inst:elsauce}}
          \and 
          N. Grieves\inst{\ref{inst:unige}}
          \and
          N.~M.~Guerrero\orcidlink{0000-0002-5169-9427}\inst{\ref{inst:florida}}
          \and 
          C.~Hellier\orcidlink{0000-0002-3439-1439}\inst{\ref{inst:keele}}
          \and 
        M.~I.~Jones\inst{\ref{inst:eso-santiago}}
       \and 
       D.~W.~Latham\orcidlink{0000-0001-9911-7388}\inst{\ref{inst:cfa}}
          \and 
                N. Law\inst{\ref{inst:ncarolina}}
                \and 
                A. W. Mann\orcidlink{[0000-0003-3654-1602}\inst{\ref{inst:ncarolina}}
                \and 
                M. Marmier\inst{\ref{inst:unige}}
                \and 
                G. Ottoni\inst{\ref{inst:unige}}
                \and 
                D.~J.~Radford\orcidlink{0000-0002-3940-2360}\inst{\ref{inst:aavso}}
                \and 
                N.~Restori\inst{\ref{inst:unige}}
                \and 
                A.~Rudat\inst{\ref{inst:kavli-mit}}
                \and
                L. Dos Santos\inst{\ref{inst:unige}, \ref{inst:stsci}}
                \and
                S. Seager\orcidlink{0000-0002-6892-6948}\inst{\ref{inst:mit_deaps}, \ref{inst:kavli-mit}, \ref{inst:mit_daa}}
                \and
                K. Stassun\orcidlink{0000-0002-3481-9052}\inst{\ref{inst:vanderbilt}}
                \and 
                C. Stockdale\orcidlink{0000-0003-2163-1437}\inst{\ref{inst:hazelwood}}
                \and 
                S. Udry\orcidlink{0000-0001-7576-6236}\inst{\ref{inst:unige}}
                \and 
                S.~Wang\orcidlink{0000-0002-7846-6981}\inst{\ref{inst:uindiana}}
                \and 
                C.~Ziegler\orcidlink{0000-0002-0619-7639}\inst{\ref{inst:dunlap}}
          }

   \institute{Départment d’astronomie de l’Université de Genève, Chemin 
   Pegasi 51, 1290 Versoix, Switzerland\label{inst:unige}\\
                 \email{janis.hagelberg@unige.ch}
             \and
             European Southern Observatory, Karl-Schwarzschild-Stra{\ss}e 2, 85748 Garching bei M{\"u}nchen, Germany\label{inst:eso-garching}
                        \and
                        Steward Observatory, University of Arizona, Tucson, AZ 85721, USA\label{inst:arizona}
                    \and 
            Department of Astrophysical Sciences, Princeton University, Princeton, NJ 08544, USA\label{inst:princeton}
            \and
            Cahill Center for Astrophysics, California Institute of Technology, Pasadena, CA 91125, USA\label{inst:caltech}
            \and
            Cerro Tololo Inter-American Observatory, Casilla 603, La Serena, Chile\label{inst:ctio}
            \and 
            Center for Astrophysics \textbar \ Harvard \& Smithsonian, 60 Garden Street, Cambridge, MA 02138, USA\label{inst:cfa}
            \and 
            Department of Astronomy, Yale University, New Haven, CT 06511, USA\label{inst:yale}
            \and
            El Sauce Observatory, Coquimbo Province, Chile\label{inst:elsauce}
            \and 
                        Bryant Space Science Center, Department of Astronomy, University of Florida, Gainesville, FL 32611,
                        USA\label{inst:florida}
            \and 
            Astrophysics Group, Keele University, Staffordshire, ST5 5BG, UK\label{inst:keele}
            \and 
            European Southern Observatory, Alonso de C\'ordova 3107, Vitacura, Casilla 19001, Santiago, Chile\label{inst:eso-santiago}
            \and 
            Department of Physics and Astronomy, The University of North Carolina at Chapel Hill, Chapel Hill, NC 27599-3255, 
            USA\label{inst:ncarolina}
            \and
            Brierfield Observatory, New South Wales, Australia\label{inst:aavso}
            \and 
                        Space Telescope Science Institute, 3700 San Martin Drive, Baltimore, MD, 21218, USA\label{inst:stsci}
            \and 
            Department of Earth, Atmospheric, and Planetary Sciences, Massachusetts Institute of Technology, Cambridge, MA 02139, 
            USA\label{inst:mit_deaps}
            \and 
            Department of Physics and Kavli Institute for Astrophysics and Space Research, Massachusetts Institute of Technology, 
            Cambridge, MA 02139, USA\label{inst:kavli-mit}
            \and
            Department of Aeronautics and Astronautics, Massachusetts Institute of Technology, Cambridge, MA 02139, USA\label{inst:mit_daa}
            \and
            Department of Physics and Astronomy, Vanderbilt University, Nashville, TN 37235, USA\label{inst:vanderbilt}
            \and
            Hazelwood Observatory, Australia\label{inst:hazelwood}
            \and
            Department of Astronomy, Indiana University, Bloomington, IN 47405, USA\label{inst:uindiana}
            \and 
            Dunlap Institute for Astronomy and Astrophysics, University of Toronto, 50 St. George Street, Toronto, Ontario M5S 3H4, 
            Canada\label{inst:dunlap} 
                          }

   \date{Received September 15, 1996; accepted March 16, 1997}

  \abstract
   {We report the discovery of a hot Jupiter on a 
   3.28-day orbit around a 1.08 \msun\ G0 star that is the secondary component in a loose binary system. 
        Based on follow-up radial velocity observations of \Nstar\ with CORALIE 
         on the Swiss 1.2\,m telescope and CHIRON on the 1.5\,m telescope at the Cerro Tololo Inter-American Observatory (CTIO), 
         we measured the planet 
        mass to be \mplanet. Two transits were further observed with CORALIE to
         determine the alignment of \Nplanet\ with respect to 
        its host star. Analysis of the Rossiter-McLaughlin signal from the planet
         shows that the sky-projected obliquity is $\lambda$ = 
        99.3$\stackrel{+3.8}{_{-3.7}}^{\circ}$. Numerical simulations show that
         the neighbour star \NstarB\ is too distant to have trapped 
        the planet in a Kozai--Lidov resonance, suggesting a different dynamical
         evolution or a primordial origin to explain this 
        misalignment. The 1.15 \msun\ primary F9 star of the system (TYC 8501-01597-1, at $\rho\sim$11\arcsec) was also
         observed with CORALIE in order to provide upper limits for 
        the presence of a planetary companion orbiting that star.   
}


\maketitle

\section{Introduction}

The thousands of exoplanets that have already been discovered show not only a wide variety in size, density, interior, and atmospheric structure but also in orbital configuration.
With every survey based on new or improved detection and characterisation techniques, this variety grows \citep[see reviews in][and references therein]{udry_statistical_2007, bowler_imaging_2016, zhu_exoplanet_2021}.

Such surveys have made it possible to highlight the strong connection between a host star's properties and those of its orbiting
 planets, which are often directly linked to their formation history, such as metallicity favouring giant planet formation
 \citep[e.g.,][]{santos_spectroscopic_2004, fischer_planet-metallicity_2005}.
These surveys also revealed that small-planet occurrence is not affected by the stellar host properties 
\citep{sousa_spectroscopic_2008, buchhave_abundance_2012} or the fact that late type 
stars tend to host smaller planets \citep{bonfils_harps_2013, mulders_stellar-mass-dependent_2015}.

Another important link has been shown with the presence of a stellar companion, such as the suppression of planet formation in close binaries, with a limit
found to be at 47~au or 58~au by \citet{kraus_impact_2016} and \citet{ziegler_soar_2021}, respectively. Indeed, \citet{Hirsch21} reported a strong drop in planet occurrence rate at binary separations of 100 au, with planet occurrence rates of $\sim20\%$ for binary separations larger than 100 au and of 4\% within 100 au. Moreover,
an overabundance, by a factor of about three, of hot Jupiters has also been noted when a stellar companion was found \citep{law_robotic_2014, ngo_friends_2016, Wang2017}, as well as an increase in eccentricities \citep{moutou_eccentricity_2017}. To a wider extent, \citet{ fontanive_census_2021} have shown
that the influence of a stellar companion is strongest for high mass planets and short orbital periods.
However, these demographic results are mainly inferred from the planetary parameters accessible through transit and radial velocity 
(RV) observations, such as the orbital periods, masses, radii, eccentricities, stellar host properties, and dynamics in the case of 
systems with multiple planets and even multiple stars.

An important additional parameter that is difficult to obtain is the orbital obliquity, or spin-orbit angle, which is the angle between the 
stellar spin axis and the axis of the orbital plane. Various approaches have been developed to
        measure it, namely, the analysis of disc-integrated RVs \citep{Queloz2010}, Doppler tomography \citep{cameron2010a}, and the reloaded Rossiter-McLaughlin effect \citep{Rossiter1924, McLaughlin1924, Cegla2016}, which requires performing high-resolution spectroscopy during a planet's transit. 

The spin-orbit angle is a tracer of a planet's history, shedding light on its formation and evolution (see reviews by \citealt{Triaud2018} and \citealt{albrecht_stellar_2022}). 
 Measuring the spin-orbit angle can help one distinguish between a smooth dynamical history that keeps the
        system aligned, such as migration within the protoplanetary disc (e.g., \citealt{Winn2015}), and
        more disruptive scenarios in which the planetary orbit is tilted through gravitational interactions
        with the star or with outer companions (e.g., \citealt{Fabrycky2007}, \citealt{Teyssandier2013}). Misaligned spin-orbit angles can 
        have either a primordial (i.e., occurring during the star-planet formation) or a post-formation origin (taking place after disc 
        dispersal). Primordial mechanisms, such as magnetic warping or chaotic accretion, tend to produce mild misalignments 
        \citep{albrecht_stellar_2022}, unless a distant stellar companion exists, which can either enhance the magnetic warping of the 
        inner disc \citep{Foucart11} or gravitationally tilt the disc \citep{Batygin12}. While the latter process has recently 
        been disfavoured due to planet-star coupling damping the effect \citep{Zanazzi2018}, the magnetic warping aided by a stellar companion 
        could lead to a large distribution of obliquities, including retrograde planets \citep{Foucart11}. 
 Post-formation misalignments are driven by gravitational interactions involving a third body (planet or star). While planet-planet scattering can produce misalignments of up to 60 degrees \citep{Chatterjee2008}, a close encounter with another star would yield an isotropic distribution of obliquities, although this process is only expected in a very dense cluster environment \citep{Hamers17}. The post-formation gravitational interactions can also lead to high-eccentricity tidal migration, where a planet that is originally far from its central star acquires high eccentricity, and its orbit is later circularised (and shrunk) due to the strong stellar tides suffered near periastron. In this process, the driver of the increase in eccentricity is the Kozai--Lidov effect induced by a third massive external companion, such as a star or a brown dwarf \citep{Kozai62}. This mechanism has been proposed as a source of hot Jupiters \citep{Dawson18} and can also lead to spin-orbit misalignment \citep{Fabrycky2007}. Recently, \citet{Vick23} showed that if the system already has a primordial misalignment (from companion-disc interaction, stellar spin-disc interaction, and disc dispersal),
 Kozai--Lidov oscillations lead to predominantly retrograde stellar obliquities, with 
 the distribution of angles peaking at a misalignment of around 90 degrees (i.e., polar orbits) for hot Jupiters.

In this work, we focused our exoplanet search around the wide stellar binary system TOI-858 A-B, which has an angular separation of 
$\rho\sim$11\arcsec\ 
($\sim$3000 au; \citealt{gaiaEDR3}). Wide binaries (i.e., binary separations in the range of 300-20'000 au) are common in our galaxy 
\citep{offner_origin_2023}, yet their origin remains a mystery because they easily become disrupted in dense star-forming 
regions \citep{offner_origin_2023}. The most accepted formation channel  for wide binaries is during the phase of star cluster 
dissolution \citep{Kouwenhoven10, Moeckel10}.  

In this study, we apply the latest Rossiter-McLaughlin technique (i.e., Revolutions, or
        RMR; \citealt{Bourrier2021}) to two transits of \Nplanet\ observed with the CORALIE spectrograph on the 1.2m Swiss telescope
        at La Silla Observatory (Sect.~\ref{sec:RM}) in order to measure the spin-orbit angle between \Nplanet \ and \Nstar. 
 The paper is structured as follows. We first describe the discovery and follow-up observations of the 1.08 \msun\ G0 \citep{pecaut_intrinsic_2013} star \Nstar\ and its nearby
and similarly bright ($\Delta$mag$_{\tess}$ = 0.248) 1.15 \msun\ F9 stellar companion \NstarB\  (Sect.~\ref{sec:obs}). Then, we 
present the joint analysis of the acquired data leading to the confirmation of the planet (Sect.\ref{sec:obs-analysis}).
Further investigation of the orbital architecture of the \Nstar\ system along
 with the link to the nearby star \NstarB\ is given in Sect.
  \ref{sec:archi}.
Finally, the results are discussed in Sect.~\ref{sec:discuss}.

\section{Observations}\label{sec:obs}
A summary of the photometry and high-resolution spectroscopy data used in the joint analysis of \Nplanet\ can be found in Tables 
\ref{tab:photometric_obs_data} and \ref{tab:rv_obs_data}. Additionally, SOAR speckle imaging was used to rule out nearby stellar 
companions, as described in Sect. \ref{sec:soar}.

\begin{table*}
        \centering
        \caption{\label{tab:photometric_obs_data} Summary of the discovery \tess\ photometry, archival WASP-South photometry, and ground-based follow-up photometry of \Nstar. }
        
        \begin{tabular}{lcrrrr}
                \hline\hline
                \noalign{\smallskip}
                Obs date & Source & Filter &         lin. limb-dark. &         quad. limb-dark. &         Dilution factor\\
                UT & & &$u_{1}$ & $u_{2}$ &$A_D$ \\
                \noalign{\smallskip}
                \hline
                \noalign{\smallskip}
                2010-08-13 -- 22-01-2012  & WASP-South 200\,mm & R &         $0.367\pm0.038$      &     $0.287\pm0.036$      &    
                $0.646\pm0.022$      \\ \vspace{0.8mm}
                2012-09-02 -- 01-12-2014 & WASP-South  85\,mm & R  &         $0.367\pm0.038$      &     $0.287\pm0.036$      &    
                $0.646\pm0.022$      \\ \vspace{0.8mm}
                2018-09-20 -- 13-11-2018  & \tess\ 30 min FFI s3 + s4  & TESS &         $0.315\pm0.028$  &         $0.293\pm0.034$  &         
                $-0.00001\pm0.00030$ \\\vspace{0.8mm}
                2019-08-28 & Hazelwood & i'  &         $0.268 \pm 0.050$      &         $0.263\pm0.049$      &      --   \\ \vspace{0.8mm}
                2019-11-05 & Brierfield & B &         $0.657\pm0.042$      &   $0.159^{+0.040}_{-0.039}$      &  --  \\ \vspace{0.8mm}
                2019-12-05 & Evans at El Sauce & B &         $0.657\pm0.042$      &   $0.159^{+0.040}_{-0.039}$      &  --  \\
                
                2020-08-26 --   2020-11-13   & \tess\ 2' {\small SPOC} s29, s30, s31 & TESS  &         $0.315\pm0.028$  &         
                $0.293\pm0.034$  &         $-0.00001\pm0.00030$\\ 
                \noalign{\smallskip}
                \hline
                \noalign{\smallskip}                            
        \end{tabular}
\end{table*}

\begin{table}
        \centering
        \caption{\label{tab:rv_obs_data} Summary of the RV observations of \Nstar. }
        
        \begin{tabular}{lcc}
                \hline\hline
                \noalign{\smallskip}
                & CORALIE & CHIRON\\
                \noalign{\smallskip}
                \hline
                \noalign{\smallskip}
                Obs date (UT) & 2019.08.13 & 2019.08.13\\
                &  --2021.01.18 & --2019.09.02\\
                No. of observations& 33 RVs & 7 RVs \\
                Relative RV Offset (m/s)& $64361 \pm 7$ & $86^{+24}_{-23}$\\
                RV Jitter (m/s) & $3^{+12}_{-0}$ &  $55^{+24}_{-20}$\\
                \noalign{\smallskip}
                \hline
                \noalign{\smallskip}            
        \end{tabular}
\end{table}

\subsection{TESS discovery photometry}

The star \Nstar\ (TIC 198008005) was observed by the Transiting Exoplanet Survey Satellite \citep[\tess\ -][]{Ricker:2015} in Camera 3 in
 sectors 3, 4, 29, 30, and 31. For the first two sectors, 30-min cadence full-frame images (FFIs) are available. \Nplanet\ was
 identified as a TESS Object of Interest (TOI) based on the MIT Quick Look Pipeline data products \citep[QLP;][]{QLP1,QLP2}. In our joint analysis, for sectors 3 and 4, we used the extracted photometry from the QLP pipeline. During the third year of the \tess\
 mission, \Nstar\ was observed with a 2-min cadence in sectors 29, 30, and 31. The light curves used in our joint analysis from these
 sectors are  publicly available at the Simple Aperture Photometry flux with Pre-search Data Conditioning
 \citep[PDC-SAP;][]{Stumpe:2014,Stumpe:2012,2012PASP..124.1000S,Jenkins:2010} provided by the Science Processing Operations Center
 \citep[SPOC; ][]{Jenkins:2016}.  

Within the large 21\arcsec\ \tess\ pixels, \Nstar\ is completely blended with a slightly 
brighter star, \NstarB\ (\NstarBTIC). The two sources are separated by 10.7\arcsec. 
Based on the \tess\ data alone, it is not possible to determine around which of the two stars the transits are taking place. A combination of seeing-limited ground-based photometry and RV measurements helped determine which of the two stars host the planet (discussed in Sec. \ref{sec:goundbasedphot} and \ref{sec:spec}).


\begin{figure}
    \centering
    \includegraphics[height=0.73\textheight]{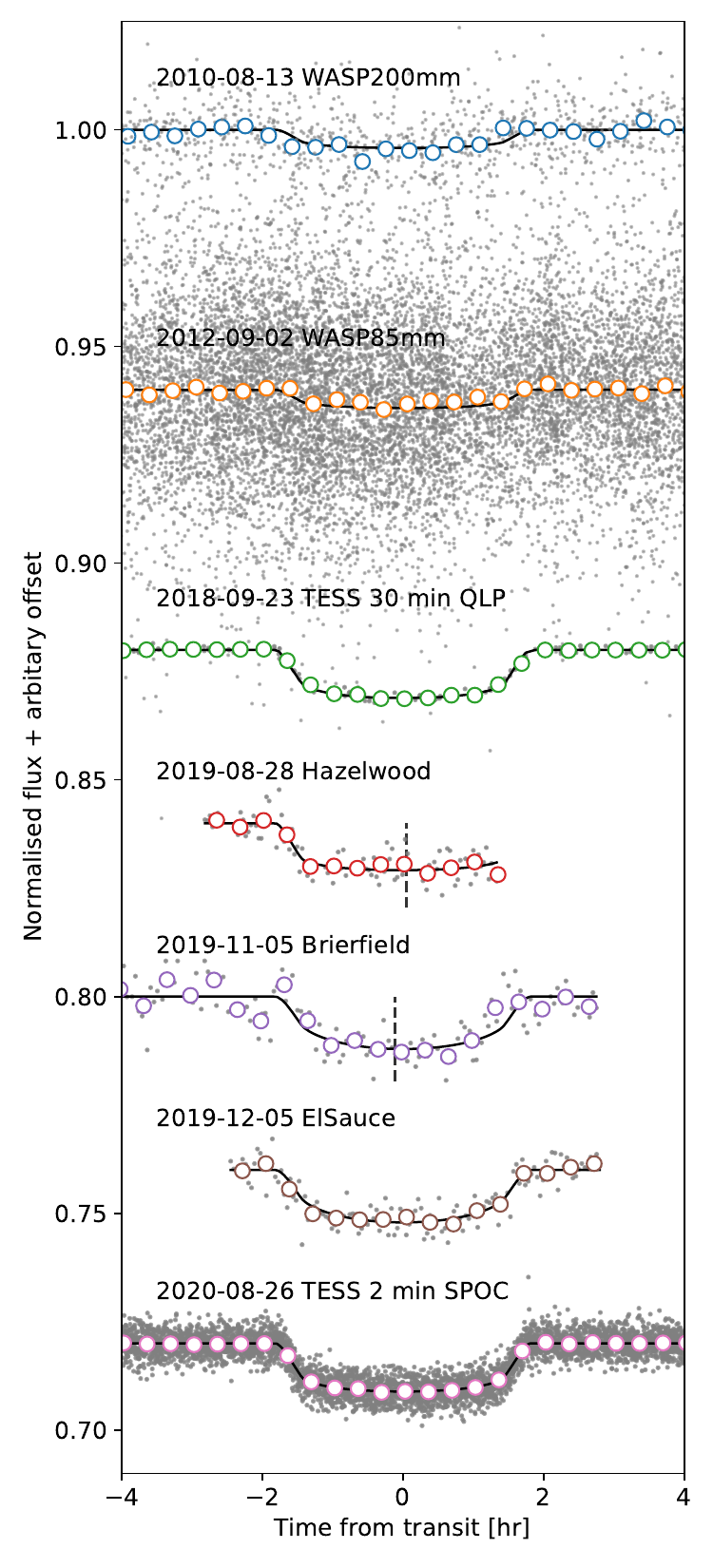}
    \includegraphics[height=0.22\textheight]{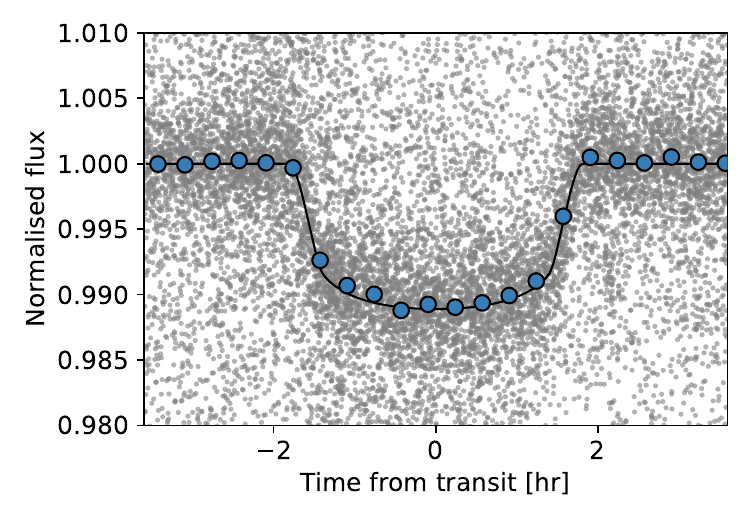}
    \caption{Photometric transit observations of \Nplanet\ in 20 min bins. The vertical dashed
    lines represent meridian flips. The bottom panel shows all light curves phase folded and overplotted.}
        \label{fig:lightcurves}%
\end{figure}

\subsection{Ground-based follow-up photometry} \label{sec:goundbasedphot}

The \textit{TESS} pixel scale is $\sim21\arcsec$ pixel$^{-1}$, and photometric apertures typically extend out to roughly 1 arcminute, 
generally causing multiple stars to blend in the \textit{TESS} photometric aperture. To determine the true source of the 
\textit{TESS} detection, we acquired ground-based time-series follow-up photometry of the field around \Nstar\ as part of the 
\textit{TESS} Follow-up Observing Program \citep[TFOP;][]{collins:2018}.\footnote{https://tess.mit.edu/followup} The follow-up light 
curves were also used to confirm the transit depth and thus the \textit{TESS} photometric de-blending factor as well as to refine the \textit{TESS} 
ephemeris and place constraints on transit depth differences across optical filter bands. We used the {\tt TESS Transit Finder}, 
which is a customised version of the {\tt Tapir} software package \citep{Jensen:2013}, to schedule our transit observations. The 
photometric data were extracted using the {\tt AstroImageJ} ({\tt AIJ}) software package \citep{Collins:2017}. The observations are 
summarised in the following subsections and in Table \ref{table:tfopobs}. As shown in Fig. \ref{fig:lightcurves}, 
they confirm that the \tess-detected transit-like event is occurring around \Nstar.

\subsubsection{Hazelwood Observatory}

We observed an ingress of \Nplanet\ in Sloan $i'$-band on UTC 2019 August 28 from Hazelwood Observatory near Churchill, Victoria, 
Australia. The 0.32\,m telescope is equipped with a $2184\times1472$ SBIG STT3200 camera. The image scale is 0$\farcs$55 
pixel$^{-1}$, resulting in a $20\arcmin\times14\arcmin$ field of view. The photometric data were extracted using a circular 
$5\farcs$9 photometric aperture.

\subsubsection{Brierfield Observatory}

We observed a full transit in B-band on UTC 2019 November 5 from Brierfield Observatory near Bellingen, New South Wales, Australia. 
The 0.36\,m telescope is equipped with a $4096\times4096$ Moravian 16803 camera. The image scale after binning 2$\times$2 is 
1$\farcs$47 pixel$^{-1}$, resulting in a $50\arcmin\times50\arcmin$ field of view. The photometric data were extracted using a 
circular $4\farcs4$ photometric aperture.

\subsubsection{Evans 0.36\,m Telescope}

We observed a full transit in B-band on UTC 2019 December 5 from the Evans 0.36\,m telescope at El Sauce Observatory in Coquimbo 
Province, Chile. The telescope is equipped with a $1536\times1024$ SBIG STT-1603-3 camera. The image scale after binning 2$\times$2 
is 1$\farcs$47 pixel$^{-1}$, resulting in an $18.8\arcmin\times12.5\arcmin$ field of view. The photometric data were extracted using 
a circular $5\farcs$9 photometric aperture.

\begin{table*}[h]
        \caption{TFOP photometric follow-up observation log.}\label{table:tfopobs}.
        \tabcolsep 1.5 mm
        \begin{tabular}{lcccccccccc}
                \hline\hline
                \noalign{\smallskip}
                Observatory & Aperture  & Filter & Date & Start & End   & Length & Exp. Time & Airmass & Comp. & Precision\\
                & (m)       &        & (UTC)& (UTC) & (UTC) & (min.)   & (sec.)    & Range   & Stars (n)  & (ppt/10 min) \\
                \noalign{\smallskip}
                \hline
                \noalign{\smallskip}
                Hazelwood   & 0.32 & $i'$  & 2019-08-28 & 15:23 & 19:36 & 253      & 120       & 1.45 - 1.04        & 3 & 1.4 \\
                Brierfield  & 0.36  & B    & 2019-11-05 & 11:07 & 18:01 & 414      & 180       & 1.46 - 1.10 - 1.33 & 4 & 2.6 \\
                Evans       & 0.36  & B    & 2019-12-05 & 01:08 & 06:28 & 320      & 150       & 1.26 - 1.09 - 1.25 & 4 & 1.3 \\
                \hline
        \end{tabular}
\end{table*}

\subsection{WASP-South archival photometry}
The \Nstar\ system was observed in 2010 and 2011 by the WASP-South survey when it was equipped with 200-mm, f/1.8 lenses observing with a 400 -- 700 nm passband \citep{2006PASP..118.1407P}. The 48-arcsec extraction aperture encompasses both stars of the binary. A total of 6445 photometric data points were obtained, spanning 110 nights starting in 2010 August and then another 170 nights starting 2011 August. The standard WASP transit-search algorithms \citep{2006MNRAS.373..799C} detect the 3.27-d periodicity (though the object was never selected as a WASP candidate) and report an ephemeris of JD(TDB) = 245\,5982.41565(13) + $E$ $\times$ 3.279765(13).

\subsection{High-resolution follow-up spectroscopy}
\label{sec:spec}

The high-resolution spectrographs CORALIE, HARPS, and CHIRON were utilised to fully characterise the \Nstar\ system. Multi-epoch monitoring allowed us to measure the reflex motion of the star(s) induced by the planet. Furthermore, we observed two spectroscopic transits of \Nplanet\ with the aim of determining the spin-orbit angle of \Nplanet.

\subsubsection{CORALIE}
Both \Nstar\ and \NstarB\ were observed with the CORALIE spectrograph mounted on the Swiss $1.2\, \mathrm{m}$ Euler telescope at 
\LSO, Chile \citep{CORALIE}. CORALIE has a resolution of $R=60,000$ and is fed by a 2\arcsec\ on-sky A fibre. An additional B fibre 
can be used to either provide simultaneous Fabry-P\'erot (FP) RV drift monitoring or on-sky monitoring of the 
background contamination. 

We obtained 22 spectra between 2019 August13 and 2021 January 18 UT with simultaneous FP to monitor the RV of \Nstar\ as it was orbited by 
\Nplanet. The exposure times ranged between 1800 and 1200\,s, depending on the observing schedule, resulting in an average S/N per pixel 
of 15 
at 5500\,\AA. 

Two spectroscopic transits were also observed on 2019 December 5 and 2021 January 18 UT. On both nights, we obtained 11 spectra with individual exposure times of 1800\,s. During the first visit (2019-12-05) one spectrum was obtained before transit, seven spectra during transit, and three after transit. We observed without any simultaneous wavelength calibration in order to enable correction for sky contamination monitored with the B fibre. In this mode, the wavelength solution originates from the calibration acquired during daytime. Instrumental drift during the night is not taken into account but were expected to be smaller than or on-par with the expected RV uncertainty for this target. After receiving ambiguous results, we repeated the observations on the second night (2021-01-18) with simultaneous FP. During this visit, we took one spectrum before transit, six spectra during transit, and four after transit.

For \NstarB, we acquired nine CORALIE spectra between 2019 October 17 and 2020 March 12 UT in order to check for additional stars and giant planets in 
the system. All spectra were taken with simultaneous FP, and nearly all had exposure times of 1800\,s, while one was set to 1200\,s. The
average S/N per pixel were 19 at 5500\,\AA.

All spectra were reduced with the standard CORALIE Data Reduction Software (DRS). Spectra for both stars were cross-correlated with a 
G2 binary mask \citep{baranne1996} to extract RV measurements as well as cross-correlation function (CCF) line diagnostics, such as 
bisector-span \citep{Queloz2001} and full width at half maximum (FWHM). We also derived H-$\alpha$ activity indicators for each 
spectrum. Two spectra taken on 2019 September 29 and 2019 September 30 UT were rejected by the automatic quality control of the DRS due to a large 
instrument drift of more than 150\,\ms, which can lead to less precise drift correction of a few \ms. For the analysis of \Nplanet, 
we deemed these drift-corrected RVs to still be useful, as the obtained RV uncertainties are larger than the error on the drift 
correction. Furthermore, the measured RV semi-amplitude is more than 100\,\ms\ (Tab. \ref{tab:rv_coralie_toi858} and 
\ref{tab:rv_coralie_tyc}).

\subsubsection{HARPS}
With the HARPS spectrograph on the 3.6\,m ESO telescope at the \LSO, Chile \citep{HARPS}, half a spectroscopic transit 
was observed on 2019 December 5. This took place during technical time, when the new HARPS+NIRPS front end \citep{NIRPS} was being 
commissioned. In total, six spectra were obtained in high efficiency mode (EGGS), which trades spectral resolution for up to twice the 
throughput by using a slightly larger on-sky fibre (1.4\,\arcsec) than the high accuracy mode (HAM, 1\arcsec) but is still small enough 
to prevent contamination from the secondary star. The first two spectra have exposure times of 900\,s, which was then decreased to 
600\,s to get a better time resolution during transit. The spectra have S/N 40-30 per pixel at 5500\,\AA, and RV uncertainties of 
3.5 -- 5\,\ms. Only one spectrum was taken out of transit.  

All HARPS spectra were reduced with the offline HARPS DRS hosted at the Geneva Observatory. Using a sufficiently wide velocity window of 60\,\kms, CCFs were derived with a G2 binary mask.

\subsubsection{CHIRON}

We obtained spectroscopic data of \Nstar\ using the CHIRON spectrograph \citep{chiron}, a high-resolution fibre-fed spectrograph 
mounted on the 1.5\,m telescope at the Cerro Tololo Inter-American Observatory (CTIO) in Chile. We obtained in total seven different 
spectra between 2019 August 9 and 2019 September 1. For these observations, we used the image slicer mode (resolving power $\sim$ 
80,000) and an exposure time of 600\,s, leading to a relatively low S/N per pixel of $\sim$ 8-10 at 5500\,\AA\ and RV uncertainties of 
20-40\,\ms. In addition, a ThAr lamp was taken before each observation, from which a new wavelength solution was computed. 
The data were reduced using the Yale pipeline, and the RVs were computed following the method described in \citet{Jones2019}, which 
has shown a long-term RV precision on $\tau$ Ceti of $\sim$ 10-15 m\,s$^{-1}$. We note that our method computes relative RVs with respect to a template that is built by stacking all of the individual spectra. Therefore, the systemic velocity is not included in the final RVs, which explains the large offset between the CHIRON and CORALIE velocities.
The Barycentric Julian Date (BJD), RV, and the corresponding 1$\sigma$ RV uncertainties are listed in Table \ref{tab:rv_chiron_toi858}.

\subsection{Speckle imaging} \label{sec:soar}

\begin{figure}
    \centering
    \includegraphics[width=\hsize]{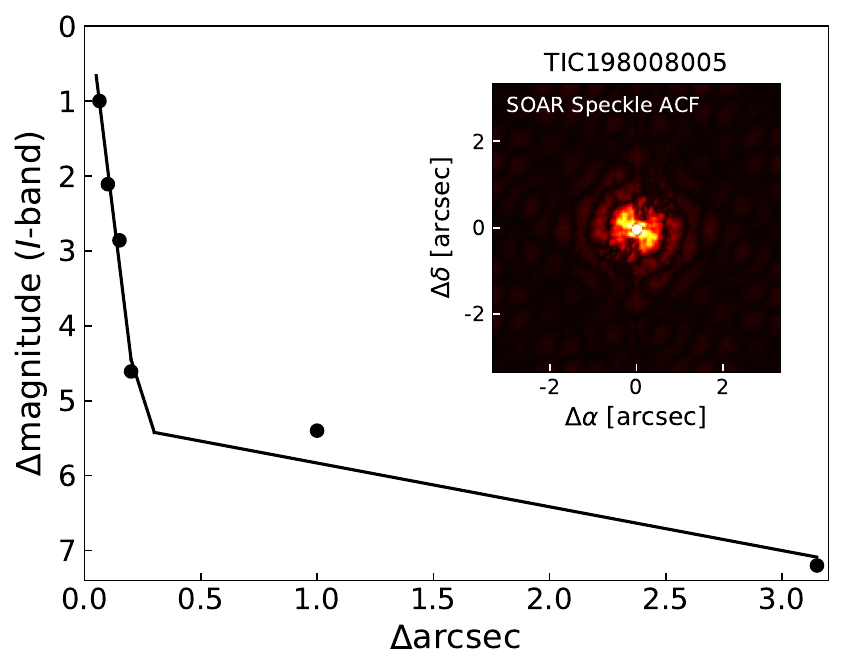}
    \caption{SOAR speckle imaging in Cousins I-band excluding nearby stars down to $\Delta$ Imag $\sim$ 5 within 3\arcsec\ of \Nstar. The inset is the speckle ACF centred on the target star.}
        \label{fig:speckle_image}%
\end{figure}

Additional nearby companion stars not previously detected in seeing-limited imaging can result in photometric contamination, reducing 
the apparent transit depth. We searched for nearby sources of \Nstar\ with SOAR speckle imaging \citep{tokovinin_ten_2018} on 2019 August 
8 UT in I-band, a visible bandpass similar to \tess. Further details of the observations from the SOAR TESS survey are available in 
\citet{ziegler_soar_2020}. We detected no nearby stars within 3\arcsec of \Nstar\ within the $5\sigma$ detection sensitivity of the 
observation, which is plotted along with the speckle auto-correlation function (ACF) in Fig. \ref{fig:speckle_image}.

\section{Analysis}\label{sec:obs-analysis}

\subsection{Spectral analysis}\label{sec:specmatch}

Stellar atmospheric parameters for \Nstar\ and \NstarB\ were derived using \emp\ \citep{specmatch}. For \Nstar,\ we stacked the six HARPS-EGGS spectra to get one high-fidelity spectrum for the analysis, and for \NstarB,\ we ran \emp\ on the stacked spectrum created from nine CORALIE spectra.

The \emp\ tool matches the input spectrum to a large spectral library of stars with well-determined parameters that have been derived with 
interferometry, optical and near-infrared (NIR) photometry, asteroseismology, and local thermodynamic equilibrium (LTE) analysis of 
high-resolution optical spectra. 
The wavelength range encompassing the Mg I b triplet (5100 - 5340 {\AA}) was utilised to match our spectra to the built-in \emp\ spectral library through $\chi^2$ minimisation. 
A weighted linear combination of the five best matching spectra were used to extract \teff, $R_{\mathrm{s}}$, and \feh. For \Nstar,\ we obtained \teff\ of $5948\pm110\mathrm{K}$, $R_{\mathrm{s}} = 1.27\pm 0.18 \rsun$, and \feh$ = 0.17\pm0.09$ (dex). 
For \NstarB,\ the spectral analysis yielded \teff $= 5911\pm110\mathrm{K}$, $R_{\mathrm{s}} = 1.47\pm       0.18 \rsun,$ and \feh $= 0.21\pm0.09$ (dex). 
The \teff\ and \feh\ were used as priors in the joint analysis detailed in Sect. \ref{sec:exofast} and \ref{sec:TIC}, which models the system using broadband photometry, GAIA information, stellar evolutionary models, and (when available) transit light curves. The final stellar parameters are listed in Tabs. \ref{tab:starB} and \ref{tab:starA}.

\subsection{Stellar rotation} \label{sec:rot}

The projected rotational velocity, $v \sin i$, was computed for each star using the calibration between $v \sin i$ and the width of the CORALIE CCF. This calibration was first presented in \cite{Santos2002}, and it has since been updated as CORALIE has undergone several updates \citep{thesis_MR}. For \Nstar,\ we obtained $v \sin i = 5.80 \pm 0.25 $ \kms, and for \NstarB,\ we obtained $v \sin i = 6.40 \pm 0.25 $ \kms. Using the stellar radii listed in Tables \ref{tab:tableTOI858} and \ref{tab:starA}, the projected rotational velocities correspond to $P_{\mathrm{rot}}/\sin i$ of $11.5\pm 0.7$\,days and $10.8\pm0.7$\,days, respectively.

\begin{figure}
    \centering
    \includegraphics[width=\columnwidth,trim={0cm 0cm 1cm 1cm},clip]{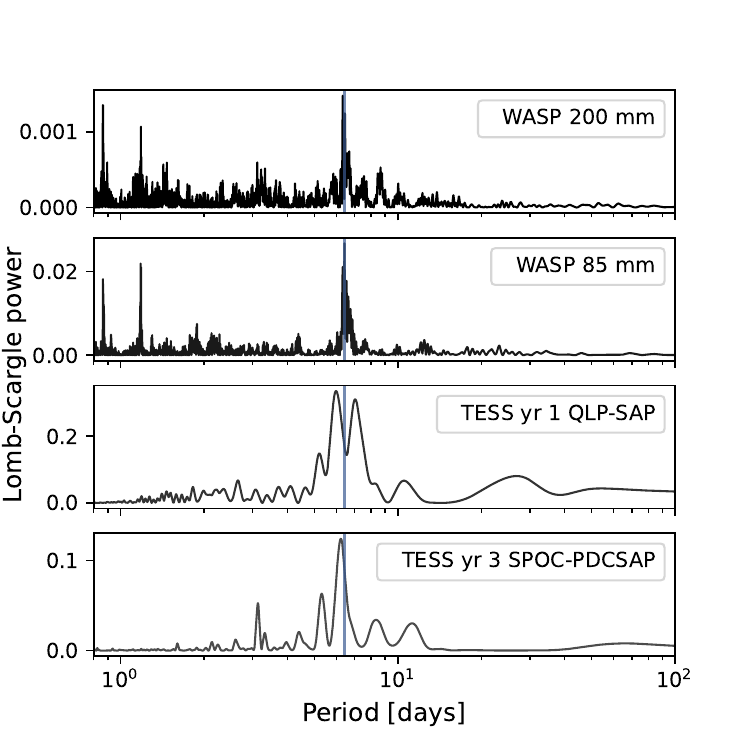}
    \caption{Periodograms showing the stellar rotational modulation in the WASP-South and TESS data, spanning 11 years in total. All data sets indicate a rotation period of 6.4 to 6.2 days. The vertical blue lines in all four panels indicate $P_{\mathrm{rot}}=6.42$\,d, determined using the WASP-South data (200\,mm and 85\,mm).}
        \label{fig:rot}%
\end{figure}

We observed clear rotational modulation in both the \tess\ and WASP-South light curves. From top to bottom, Figure \ref{fig:rot} 
shows the Lomb-Scargle 
periodograms computed for the WASP-South data before and after the change to the 85\,mm lenses in 2010 and 2011, the TESS QLP light curve 
without de-trending (SAP), and the SPOC-de-trended (PDCSAP) light curve. The transits of \Nplanet\ were masked to 
avoid picking up the planetary signal. A clear and persistent modulation can be seen at a period of 6.2 to 6.4 days in both the 
WASP-South and \tess\ data, having an amplitude varying between 2 and 3 mmags. The TESS light curves cover only a few stellar 
rotations, whereas the two WASP-South data sets each cover multiple seasons. 

Using the WASP-South data only and a modified Lomb-Scargle periodogram approach that is tailored to the noise characteristics of WASP data, as discussed in \citet{2011PASP..123..547M}, we found a rotation period of $P_{\mathrm{rot}} = 6.42 \pm 0.10$\,d. Since \Nstar\ and \NstarB\ are fully blended in both WASP-South and \tess, we could not determine which of the stars is responsible for the rotational modulation. There are no signs of two distinct rotational signals in the photometry nor at the $P_{\mathrm{rot}}/\sin i$ derived from the CORALIE CCFs. For \Nstar,\ an inclination of $34^{\circ}$ is needed to align the $P_{\mathrm{rot}}$ measured from the light curves with the spectroscopic $P_{\mathrm{rot}}/\sin i$, and for \NstarB, it is $36^{\circ}$.

For $P_{\mathrm{rot}} = 6.42 \pm 0.02$\,d, gyro-chronology yields an age of $0.3 - 0.4\,\mathrm{Gyr}$ \citep{Barnes2007} for either
 of the two stars. This is not in agreement with the ages derived in Sect.~\ref{sec:exofast} based on a spectral energy distribution
 (SED) fitting with the Mesa Isochrones and Stellar Tracks (MIST) evolutionary models. Moreover, following the approach described in \cite{2021AJ....162..197B}, we found
  no sign of Li absorption in the high S/N stacked spectra for either star, which could otherwise support a hypothesis of a young age.
  The discrepancy between the stellar age derived with gyro-chronology and MIST could indicate that the planet-hosting star \Nstar\
   has been spun-up by the planet, as such is the case for HAT-P-11b \citep{2010ApJ...710.1724B}, which shows evidence of tidal
    spin and high stellar activity \citep{2017ApJ...848...58M, tejada_arevalo_further_2021}. We note that the rotation period of $6.42
    \pm 0.02$\,d is close to twice the planetary orbital period, $P_{\mathrm{b}} = 3.28$\,d. The similar $v \sin i$ measured for the
     two stars could in this case be explained by differences in inclination. 

\begin{table}
        \centering
        \caption{\label{tab:starB} Stellar properties for \Nstar.}
                \begin{tabular}{lcc}
                        \hline\hline
                        \noalign{\smallskip}
                        Property        &       Value   &       Source\\
                        \noalign{\smallskip}
                        \hline
                        \noalign{\smallskip}
                        \multicolumn{3}{l}{Other Names}\\
                        2MASS ID        &  J04004794-5435342    & 2MASS \\
                        Gaia ID & 4683737294569921664   & Gaia EDR3 \\
                        TIC  ID & 198008005 & \tess \\
                        TOI & TOI-858 & \tess \\
                        \\
                        \multicolumn{3}{l}{Astrometric Properties}\\
                        R.A.            &        04:00:47.96    & \tess \\
                        Dec                     &         -54:35:34.5   & \tess   \\
                        $\mu_{{\rm R.A.}}$ (\masy) & 11.036  $\pm$0.017 & Gaia EDR3 \\
                        $\mu_{{\rm Dec.}}$ (\masy) & -11.004 $\pm$0.018 & Gaia EDR3\\
                        RV (km s$^{-1}$)& 64.7 $\pm$ 0.7 & Gaia EDR3 \\
                        Parallax  (mas) & 3.9727 $\pm$ 0.0134 & Gaia EDR3\\
                        \\
                        \multicolumn{3}{l}{Photometric Properties}\\
                        V (mag)         & 11.18 $\pm$ 0.07      &Tycho \\
                        G (mag)         & 11.0695  $\pm$ 0.0004 &{Gaia}\\
                        T (mag)     &   10.6444 $\pm$ 0.006     &\tess\\
                        J (mag)         & 10.06  $\pm$ 0.02     &2MASS\\
                        H (mag)         &  9.79 $\pm$ 0.03      &2MASS\\
                        K$_{\rm s}$ (mag) &9.72$\pm$ 0.02       &2MASS\\
                        W1 (mag) & 9.57 $\pm$ 0.03      &WISE\\
                        W2 (mag) & 9.58 $\pm$ 0.03      &WISE\\
                        W3 (mag) & 9.57 $\pm$ 0.03 & WISE\\
                        
                        \multicolumn{3}{l}{Spectroscopic Properties}\\
                        
                        {\it v}\,sin\,{\it i} (\kms)    & $ 5.8 \pm 0.25$       & Sec. \ref{sec:rot} \\
                        
                        \noalign{\smallskip}
                        \hline
                        \noalign{\smallskip}
                \end{tabular}
        \\
        Tycho \citep{Tycho}; 2MASS \citep{2MASS}; WISE \citep{WISE}; Gaia \citep{gaia2016,gaiaEDR3} 
\end{table}

\subsection{Joint modelling of radial velocities and transit light curves} \label{sec:exofast}

The \tess\ photometry, ground-based follow-up transit photometry, WASP-South archival light curves, and RV measurements from CORALIE and CHIRON were jointly modelled using \exofast\ \citep{Eastman2013,exofastv2}. In this approach, both stellar and planetary parameters are derived for any number of transits and RV instruments while exploring the large parameter space through a differential evolution Markov chain and Metropolis-Hastings Monte Carlo sampler (MCMC).

\begin{figure}
    \centering
    \includegraphics[width=\columnwidth,trim={0cm 0cm 1cm 13cm},clip]{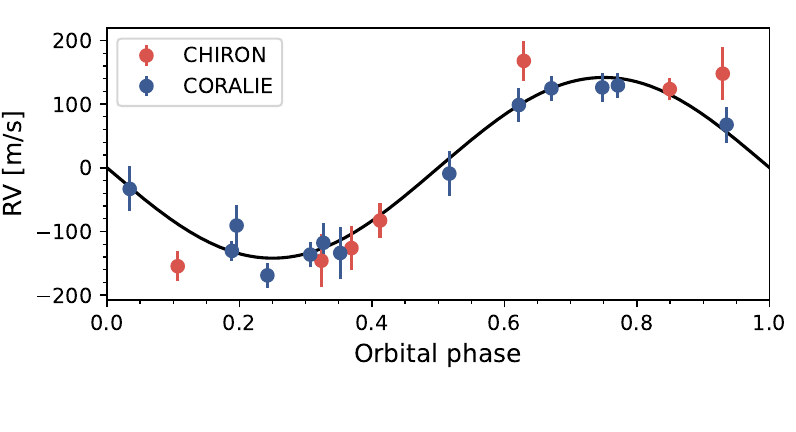}
    \caption{Phase folded CORALIE and CHIRON RV measurements for \Nstar.}
        \label{fig:RVs}%
\end{figure}

\begin{table}
        \caption{Informative priors invoked in the \exofast\ model for the \Nstar\ system. Fitted parameters not listed here use 
        uninformed, uniform priors (apart from the limb-darkening parameters, which were tabulated within \exofast\ using 
        \cite{ClaretBloemen:2011,Claret2017} and are based on the fitted stellar parameters at each step in the MCMC). \label{tab:priors}}
        \centering   
        {\small   
                \begin{tabular}{lccc}
                        \hline\hline
                        \noalign{\smallskip}
                        ~~~Parameter & Units & Prior\\
                        \noalign{\smallskip}
                        \hline
                        \noalign{\smallskip}
                        
                        ~~~~$T_{\rm eff}$\dotfill &Effective Temperature (K)\dotfill & $\mathcal{N}(5948,110)$\\
                        ~~~~$[{\rm Fe/H}]$\dotfill &Metallicity (dex)\dotfill & $\mathcal{N}(0.17,0.09)$ \\
                        ~~~~$\varpi$\dotfill &Parallax (mas)\dotfill & $\mathcal{N}(3.9727 ,0.0134 )$ \\
                        ~~~~$A_V$\dotfill &V-band extinction (mag)\dotfill & $\mathcal{U}(0, 0.04)$ \\
                        ~~~~$A_D$\dotfill &Dilution \tess \dotfill &$\mathcal{N}(0.0000,0.0003)$\\
                        ~~~~$A_D$\dotfill &Dilution WASP (R) \dotfill & $\mathcal{N}(0.55,0.05)$\\
\noalign{\smallskip}
\hline
\noalign{\smallskip}
\end{tabular}
}
\end{table}

The transit model is based on the analytical expressions in \cite{MandelAgol:2002}, and the RVs are modelled as a classic Keplerian orbit. The planet properties are described by seven free parameters: RV semi-amplitude ($K$), planet radius (\rpl), orbital inclination ($i$), orbital period ($P$), time of conjunction ($T_C$), eccentricity ($e$), and argument of periastron ($\omega_*$). Two additional RV terms, systemic velocity and RV jitter, are also fitted for each instrument (CORALIE \& CHIRON). Because the CHIRON RVs are derived with respect to a median spectral template, the systemic velocity for the instrument is therefore close to zero.

For the transit light curves, a set of two limb-darkening coefficients for each photometric band were evaluated by interpolating tables from \cite{ClaretBloemen:2011,Claret2017}. This was done within \exofast\ at each MCMC step. The limb-darkening coefficients are fitted along with the out-of-transit baseline flux and variance. The WASP-South data were heavily blended, with more than  50\% of the flux in the aperture coming from \NstarB. To take this into account, we fitted a dilution parameter to the  WASP-South light curves. A Gaussian prior on the dilution factor, based on the \gaia\ $G_{RP}$ magnitudes of the two stars, was imposed. Similarly, we also included a dilution term for the TESS data to take imperfect de-blending into account and to propagate the error that might come with it. The light curves from Hazelwood and Brierfield include meridian flips, indicated by the dashed lines
in Fig. \ref{fig:lightcurves}. Any offset between the data obtained before and after a meridian flip was modelled as a 
multiplicative de-trending term that multiplies the flux after the meridian flip with a constant. Furthermore, the El Sauce light 
curve was multiplicatively de-trended with airmass within \exofast. The Brierfield time series was likewise de-trended against total 
flux counts. (For more information on the de-trending, see Sect. 11 of \cite{exofastv2}.)

Along with the planetary properties, the stellar parameters were also modelled at each step in the MCMC. This allowed us to utilise 
the information on transit duration and orbital eccentricity embedded in the transit light curves and RVs to constrain the stellar 
density \citep{2003ApJ...585.1038S,2012MNRAS.421.1166K,2022arXiv220914301E}. We imposed Gaussian priors on \teff\ and \feh\ from the 
spectral analysis presented in Sect. \ref{sec:specmatch} while fitting the SED based on archival broadband photometry presented in 
Table \ref{tab:starB}. When including the \gaia\ DR3 parallax as a Gaussian prior, we obtained a tight constraint on the stellar 
radius. We also included an upper limit on the V-band extinction from \cite{Schlegel:1998} and \cite{Schlafly} to constrain 
line-of-sight reddening. Table \ref{tab:priors} lists the informative priors described in this section and summarises the values 
applied. To improve the stellar mass we obtained from combining the stellar radius with the stellar density from the transit light 
curve, we queried the MIST models \citep[][]{Mist0,Mist1}. This meant comparing the fitted stellar model 
parameters to viable MIST values at each step of the MCMC. The joint model was penalised for the difference between the two. This 
method uses MIST to guide the stellar parameters rather than to define them and can help break degeneracies encountered when using only 
isochrone models.
Despite \exofast\ having the ability to include Doppler tomography and the Rossiter-McLaughlin effect in its joint model, we found 
that a more sophisticated analysis of the spectroscopic transit data was needed, as outlined in the Sect. \ref{sec:RM}. 

\begin{table}
        \caption{Stellar parameters for \NstarB. \label{tab:starA}}
        \begin{tabular}{lcc} 
                \hline \hline
                \noalign{\smallskip}
                Property        &       Value   &       Source\\
                \noalign{\smallskip}
                \hline
                \noalign{\smallskip}
                \multicolumn{3}{l}{Other Names}\\
                ~~~~2MASS ID    &  J04004794-5435342  & 2MASS \\
                ~~~~Gaia ID &   4683737294570307968       & Gaia EDR3 \\
                ~~~~TIC  ID &   198008002 & \tess \\
                \\
                \multicolumn{3}{l}{Astrometric Properties}\\
                ~~~~R.A.                &               04:00:47.96 & \tess     \\
                ~~~~Dec                 &         -54:35:34.5   & \tess \\
                ~~~~$\mu_{{\rm R.A.}}$ (\masy) & 8.5707 $\pm$ 0.0187 & Gaia EDR3 \\
                ~~~~$\mu_{{\rm Dec.}}$ (\masy) & -12.6905 $\pm$ 0.0197 & Gaia EDR3\\
                ~~~~RV (km s$^{-1}$)& 65.5 $\pm$ 0.6 & Gaia EDR3 \\
                ~~~~Parallax  (mas) & 4.0181 $\pm$ 0.0146 & Gaia EDR3\\
                \\
                \multicolumn{3}{l}{Photometric Properties}\\
                ~~~~V (mag)             & 11.66 $\pm$ 0.08      &Tycho \\
                ~~~~B (mag)             & 11.07 $\pm$ 0.08      &Tycho\\
                ~~~~G (mag)             &       10.7895  $\pm$ 0.0008   &{Gaia}\\
                ~~~~T (mag)         &           10.396  $\pm$ 0.006     &\tess\\
                ~~~~J (mag)             & 9.88  $\pm$ 0.02      &2MASS\\
                ~~~~H (mag)             &  9.61 $\pm$ 0.02      &2MASS\\
                ~~~~K$_{\rm s}$ (mag) &9.56$\pm$ 0.02   &2MASS\\
                ~~~~W1 (mag) & 9.53 $\pm$ 0.03  &WISE\\
                ~~~~W2 (mag) & 9.55 $\pm$ 0.02  &WISE\\
                ~~~~W3 (mag) & 9.49 $\pm$ 0.03 & WISE\\
                ~~~~W4 (mag) & 9.35 $\pm$ 0.40 & WISE \\

                \multicolumn{3}{l}{Spectroscopic Properties}\\
                
                ~~~~{\it v}\,sin\,{\it i} (\kms) &$ 6.4 \pm 0.25$& Sec. \ref{sec:rot} \\
                ~~~~$V_{sys}$ (\kms) & 65.523 $\pm$ 0.006 & Sec. \ref{sec:TIC} \\
                \\
                \multicolumn{3}{l}{This work, joint analysis of broadband photometry, and RVs}\\
                ~~~Parameter & Units & Values\\
                \noalign{\smallskip}
                \hline
                \noalign{\smallskip}
                ~~~~$M_*$\dotfill &Mass (\msun)\dotfill &$1.152^{+0.074}_{-0.082}$\\
                ~~~~$R_*$\dotfill &Radius (\rsun)\dotfill &$1.374^{+0.040}_{-0.039}$\\
                ~~~~$L_*$\dotfill &Luminosity (\lsun)\dotfill &$2.074^{+0.061}_{-0.060}$\\
                ~~~~$\rho_*$\dotfill &Density (cgs)\dotfill &$0.623^{+0.081}_{-0.070}$\\
                ~~~~$\log{g}$\dotfill &Surface gravity (cgs)\dotfill &$4.222^{+0.042}_{-0.043}$\\
                ~~~~$T_{\rm eff}$\dotfill &Effective Temp. (K)\dotfill &$5907^{+84}_{-80}$\\
                ~~~~$[{\rm Fe/H}]$\dotfill &Metallicity (dex)\dotfill &$0.196\pm0.078$\\
                ~~~~$Age$\dotfill &Age (Gyr)\dotfill &$5.2^{+2.8}_{-2.0}$\\
                ~~~~$A_V$\dotfill &V-band extinction (mag)\dotfill &$0.023^{+0.012}_{-0.015}$\\
                ~~~~$\sigma_{SED}$\dotfill &SED error scaling \dotfill &$0.80^{+0.31}_{-0.17}$\\
                ~~~~$\varpi$\dotfill &Parallax (mas)\dotfill &$4.018\pm0.014$\\
                ~~~~$d$\dotfill &Distance (pc)\dotfill &$248.88^{+0.84}_{-0.85}$\\
    \noalign{\smallskip}
\hline
\noalign{\smallskip}
\end{tabular}\\
Tycho \citep{Tycho}; 2MASS \citep{2MASS}; WISE \citep{WISE}; Gaia \citep{gaia2016,gaiaEDR3} 
\end{table}

\begin{table*}
        \caption{Median values and 68\% confidence intervals for the \Nstar{} system. The orbital eccentricity is fixed at zero in the final 
        model, though  we list the 3$\sigma$ upper limit from models where the eccentricity was a fitted parameter. 
        \label{tab:tableTOI858}}
        \centering   
        {\small   
                \begin{tabular}{lcc lcc}
                        \hline\hline
                        \noalign{\smallskip}
                        ~~~Parameter & Units & Values & ~~~Parameter & Units & Values\\
                        \noalign{\smallskip}
                        \hline
                        \noalign{\smallskip}
                        \multicolumn{2}{l}{Stellar Parameters:}&&\multicolumn{2}{l}{Planetary Parameters:}&b\smallskip\\
                        ~~~~$M_*$\dotfill &Mass (\msun)\dotfill &$1.081^{+0.076}_{-0.070}$ & $M_P$\dotfill & Mass (\mj)\dotfill 
                        &$1.10^{+0.08}_{-0.07}$\\
                        ~~~~$R_*$\dotfill &Radius (\rsun)\dotfill &$1.308^{+0.037}_{-0.038}$ & $R_P$\dotfill &Radius (\rj)\dotfill 
                        &$1.255\pm0.039$\\
                        ~~~~$L_*$\dotfill &Luminosity (\lsun)\dotfill &$1.790^{+0.095}_{-0.083}$ & $P$\dotfill &Period (days)\dotfill 
                        &$3.2797178\pm0.0000014$\\
                        ~~~~$\rho_*$\dotfill &Density (cgs)\dotfill &$0.680^{+0.079}_{-0.064}$ & $T_C$\dotfill &Time of conjunction ({\small 
                        \bjdtdb})\dotfill &$58386.45235\pm0.00028$\\
                        ~~~~$\log{g}$\dotfill &Surface gravity (cgs)\dotfill &$4.238^{+0.038}_{-0.035}$ & $a$\dotfill &Semi-major axis 
                        (AU)\dotfill &$0.04435^{+0.0010}_{-0.00098}$\\
                        ~~~~$T_{\rm eff}$\dotfill &Effective Temperature (K)\dotfill & $5842^{+84}_{-79}$ & $i$\dotfill &Inclination 
                        (Degrees)\dotfill 
                        &$86.80^{+0.50}_{-0.44}$\\
                        ~~~~$[{\rm Fe/H}]$\dotfill &Metallicity (dex)\dotfill &$0.153\pm0.091$ & $T_{eq}$\dotfill &Equilibrium temperature 
                        (K)\dotfill &$1529^{+23}_{-22}$\\
                        ~~~~$Age$\dotfill &Age (Gyr)\dotfill &$6.8^{+2.9}_{-2.5}$ & $K$\dotfill &RV semi-amplitude (m/s)\dotfill &$143\pm 7$\\
                        ~~~~$A_V$\dotfill &V-band extinction (mag)\dotfill &$0.022^{+0.013}_{-0.014}$ & $\delta$\dotfill 
                        &$\left(R_P/R_*\right)^2$ \dotfill &$0.00974^{+0.00014}_{-0.00016}$ \\
                        ~~~~$\sigma_{SED}$\dotfill &SED photometry error scaling \dotfill &$3.10^{+1.3}_{-0.80}$ & $T_{14}$\dotfill &Total 
                        transit duration (days)\dotfill &$0.15010\pm0.00090$ \\
                        ~~~~$\varpi$\dotfill &Parallax (mas)\dotfill &$3.972\pm0.013$ & $T_{FWHM}$\dotfill &{\small FWHM} transit duration 
                        (days)\dotfill 
                        & $0.13393 \pm 0.00060 $\\
                        ~~~~$d$\dotfill &Distance (pc)\dotfill & $251.76\pm0.85$ & $b$\dotfill &Transit Impact parameter \dotfill 
                        & $0.419^{+0.050}_{-0.062}$ \\
                        &&& $e$\dotfill &Eccentricity \dotfill & < $0.15$ at 3 $\sigma$\\ 
                        &&& $\rho_P$\dotfill &Density (cgs)\dotfill &$0.690^{+0.087}_{-0.073}$\\
                        &&& $logg_P$\dotfill &Surface gravity \dotfill &$3.239^{+0.041}_{-0.039}$\\
                        &&& $\fave$\dotfill &Incident Flux (\fluxcgs)\dotfill &$1.237^{+0.076}_{-0.072}$
                        \smallskip\\
                        \noalign{\smallskip}
                        \hline
                        \noalign{\smallskip}
                \end{tabular}
        }
\end{table*}

\subsection{Stellar properties of \NstarB} \label{sec:TIC}


We derived stellar parameters for \NstarB\ in a similar way as outlined for \Nstar\ in Sect. \ref{sec:exofast}, using \exofast\ to perform an SED fit combined with MIST, while no additional information on stellar density from a transit light curve was available. 
We used Gaussian priors on \teff$=5911\pm110\,\mathrm{K}$, \feh$ =0.21\pm0.09\,\mathrm{(dex)} $ and parallax of $4.0181  \pm 0.0146\,\mathrm{(mas)}$. 
For the V-band extinction, an upper limit from dust maps of 0.04 mag was used. 
The final stellar properties of \NstarB\ are listed in Table \ref{tab:starA}. The ages we got for \Nstar\ and \NstarB\ are in agreement with each other, though poorly constrained. 

\begin{figure}
    \centering
    \includegraphics[width=\columnwidth,trim={0cm 0cm 0cm 0cm},clip]{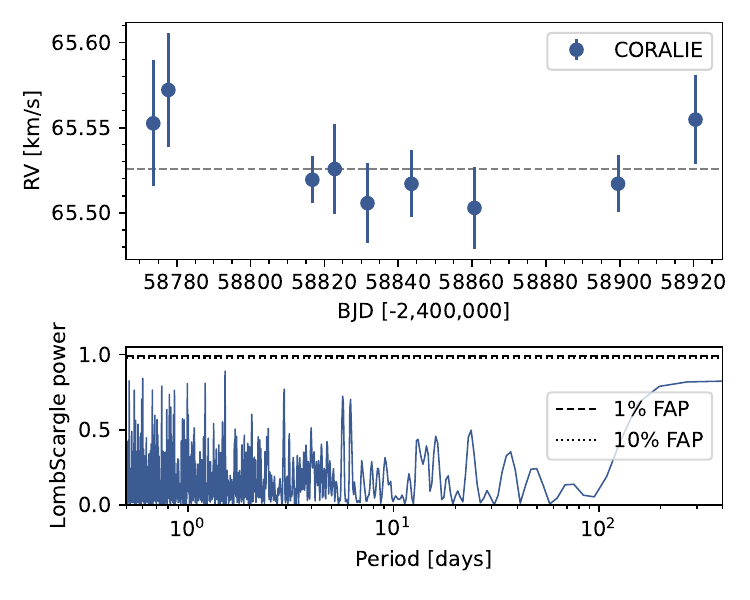}
    \caption{CORALIE RV measurements of \NstarB.
    	Top panel: RV time series showing no clear signs of a giant planet in a short period orbit. 
    Lower panel: Lomb-Scargle periodogram for the RVs with no significant signals detected. The false alarm probability (FAP) levels
     of 1\%\ and 10\% are indicated as horizontal lines.}
        \label{fig:flatRVs}%
\end{figure}

The CORALIE RVs of \NstarB\ showed no sign of planets, though only giant planets in relatively short period orbits could be ruled out. Figure \ref{fig:flatRVs} shows the RV time series with a generalised Lomb-Scargle periodogram at the bottom. No periodic signals were detected down to a false alarm probability of 10\%. We found the systemic velocity of \NstarB\ measured with CORALIE to be $65.523 \pm 0.006$\,\kms, which is in agreement with the Gaia EDR3 RV of $65.5 \pm 0.6$\,\kms and similar to the $64.7 \pm 0.7$\,\kms of \Nstar\ found by Gaia \citep{gaiaEDR3}.

\section{Analysis of \Nstar\ orbital architecture}\label{sec:archi}

We further investigated the orbital architecture of the \Nstar, \Nplanet,
and \NstarB\ ensemble. To carry out this study, we characterised the
 planet orbit through a Rossiter-McLaughlin observation (described in
 Sect. \ref{sec:RM}), and we analysed the link between the two
 stars and their possible impact on the planet orbit ( Sect. 
 \ref{sec:dynamic}). 

\subsection{Rossiter-McLaughlin Revolutions analysis}\label{sec:RM}
        
        \subsubsection{Transit observations}
        \label{sec:rm-obs}

        We utilised the two data sets obtained with CORALIE during the transit of \Nplanet\ on 2019 December 5 (Visit
        1) and 2021 January 18 (Visit 2). Spectra were extracted from the detector images and corrected and calibrated by version 3.8 of the DRS
         \citep[][]{baranne1996, queloz_fibre-fed_1999,Bouchy2001} pipeline. One of the corrections concerns the colour effect caused by
        the variability of extinction induced by Earth's atmosphere (e.g., \citealt{bourrier2014b},
        \citealt{Bourrier_2018_Nat}, \citealt{Wehbe2020}). The flux balance of the \Nstar\ spectra was reset
        to a K1 stellar spectrum template before the spectra were passed through weighted cross-correlation
        (\citealt{baranne1996}; \citealt{pepe2002}) with a G2 numerical mask to compute  the CCFs. Since the CCFs are oversampled by 
        the DRS with a step of 0.5\,km\,s$^{-1}$, for a
        pixel width of about 1.7\,km\,s$^{-1}$, we kept one in three points in all CCFs prior to their
        analysis.    
        We analysed the two CORALIE visits using the RMR technique, which follows three successive 
   steps that are described hereafter (a full description can be found in \citealt{Bourrier2021}). 
        
        \subsubsection{Extraction of the planet-occulted CCFs}
        \label{sec:step1}
        
\begin{figure}
                \includegraphics[trim=0cm 0cm 0cm 0cm,clip=true,width=\columnwidth]{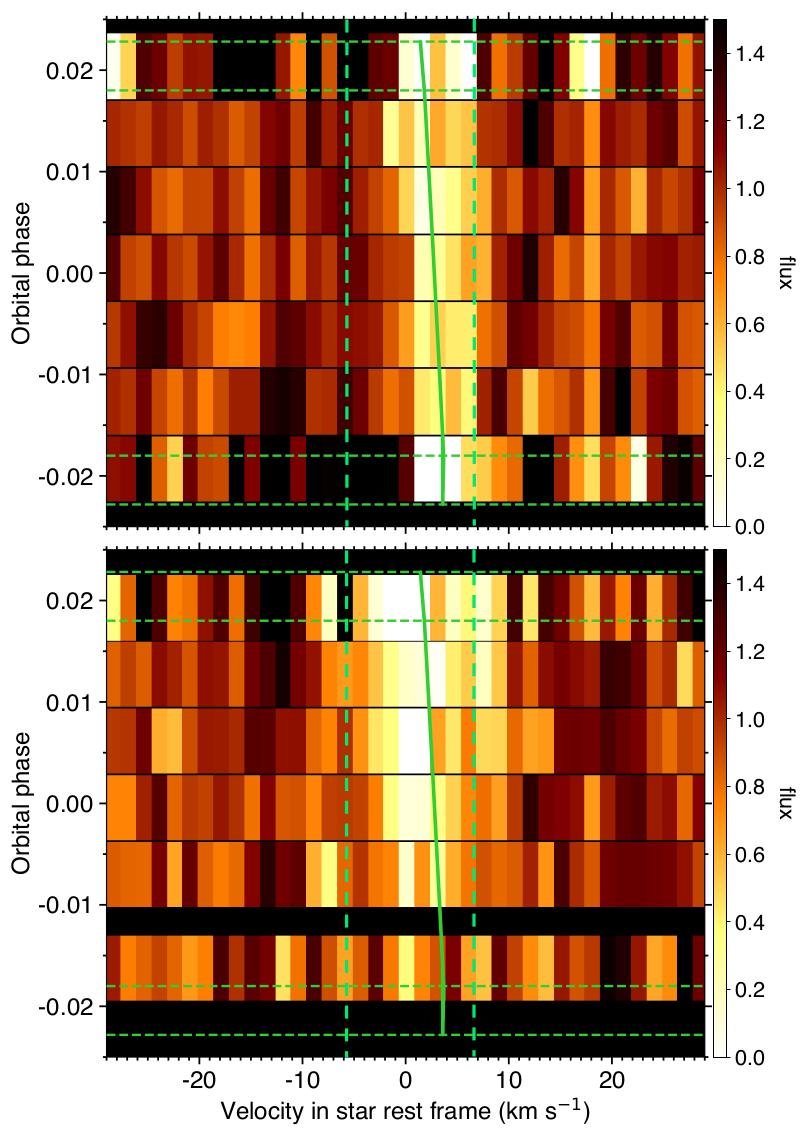}
                \centering
                \caption[]{Maps of the CCF$_\mathrm{intr}$ during the transit of \Nplanet\ in Visit 1 (upper panel)
                        and Visit 2 (lower panel). Transit contacts are shown as green horizontal dashed lines. Values are coloured as a
                        function of their normalised flux and plotted as a function of RV in the stellar rest frame (in
                        abscissa) and orbital phase (in ordinate). The stellar lines from the planet-occulted regions are
                        clearly visible in both visits. The green solid lines show the best-fit model for the stellar
                        surface RVs derived from a joint RMR fit to both data sets. The green vertical dashed lines show the spectroscopic, sky-projected stellar rotational velocity.}
                \label{fig:CCFintr_map}
        \end{figure}

        In the first step of the RMR technique, the disc-integrated CCF$_\mathrm{DI}$ were aligned by shifting their velocity table 
        with the Keplerian motion of the star, as calculated using the median values for the stellar and planet properties from the 
        joint fit analysis done in EXOFASTv2, as given in Table~\ref{tab:tableTOI858}. The continuum of the CCF$_\mathrm{DI}$ was 
        then scaled to the same flux outside of the transit and to the flux simulated during transit with the batman package
        (\citealt{Kreidberg2015}). The transit depth was taken from Table~\ref{tab:tableTOI858}, and limb-darkening
        coefficients were calculated with the EXOFAST calculator (\citealt{Eastman2013}) for the stellar
        properties listed in Table~\ref{tab:tableTOI858}. This yielded u$_1$ = 0.45 and u$_2$ = 0.27 in the visible band,
        representative of the CORALIE spectral range. The CCF$_\mathrm{DI}$ outside of the transits were co-added
        to build master-out CCFs representative of the unocculted star. Gaussian profiles were fitted to the
        master-out CCF$_\mathrm{DI}$ to determine the RV zero points in Visit 1
        (64.353$\pm$0.015\,km\,s$^{-1}$) and Visit 2 (64.358$\pm$0.013\,km\,s$^{-1}$), which were then used
        to shift all CCF$_\mathrm{DI}$ to the star rest frame. The CCFs from the planet-occulted regions were
        retrieved by subtracting the scaled CCF$_\mathrm{DI}$ from their corresponding master-out. They were
        finally normalised to a common flux level by dividing their continuum with the flux scaling applied to
        the CCF$_\mathrm{DI}$, yielding intrinsic CCF$_\mathrm{intr}$ that directly trace variations in
        the local stellar line profiles (Fig.~\ref{fig:CCFintr_map}). Flux errors were assigned to the
        CCF$_\mathrm{intr}$ as the standard deviation in their continuum flux.\\
        
        \subsubsection{Analysis of individual exposures}
        
        In the second step, a Gaussian profile was fitted to the CCF$_\mathrm{intr}$ in each exposure, over [-30 , 30]\,km\,s$^{-1}$ in the star rest frame. We sampled the posterior distributions of its RV centroid, FWHM, and contrast using \textit{emcee} MCMC (\citealt{Foreman2013}). We set uniform priors on the RV centroid with a boundary between -5 to 10\,km\,s$^{-1}$, on the FWHM with a boundary between 0 and 20\,km\,s$^{-1}$, and on the contrast (bounded between -2 and 2). One hundred walkers were run for 2000 steps, with a burn-in phase of 500 steps, to ensure that the resulting chains converged and are well mixed.  
        
\begin{figure}
                \includegraphics[trim=0cm 0cm 0cm 0cm,clip=true,width=\columnwidth]{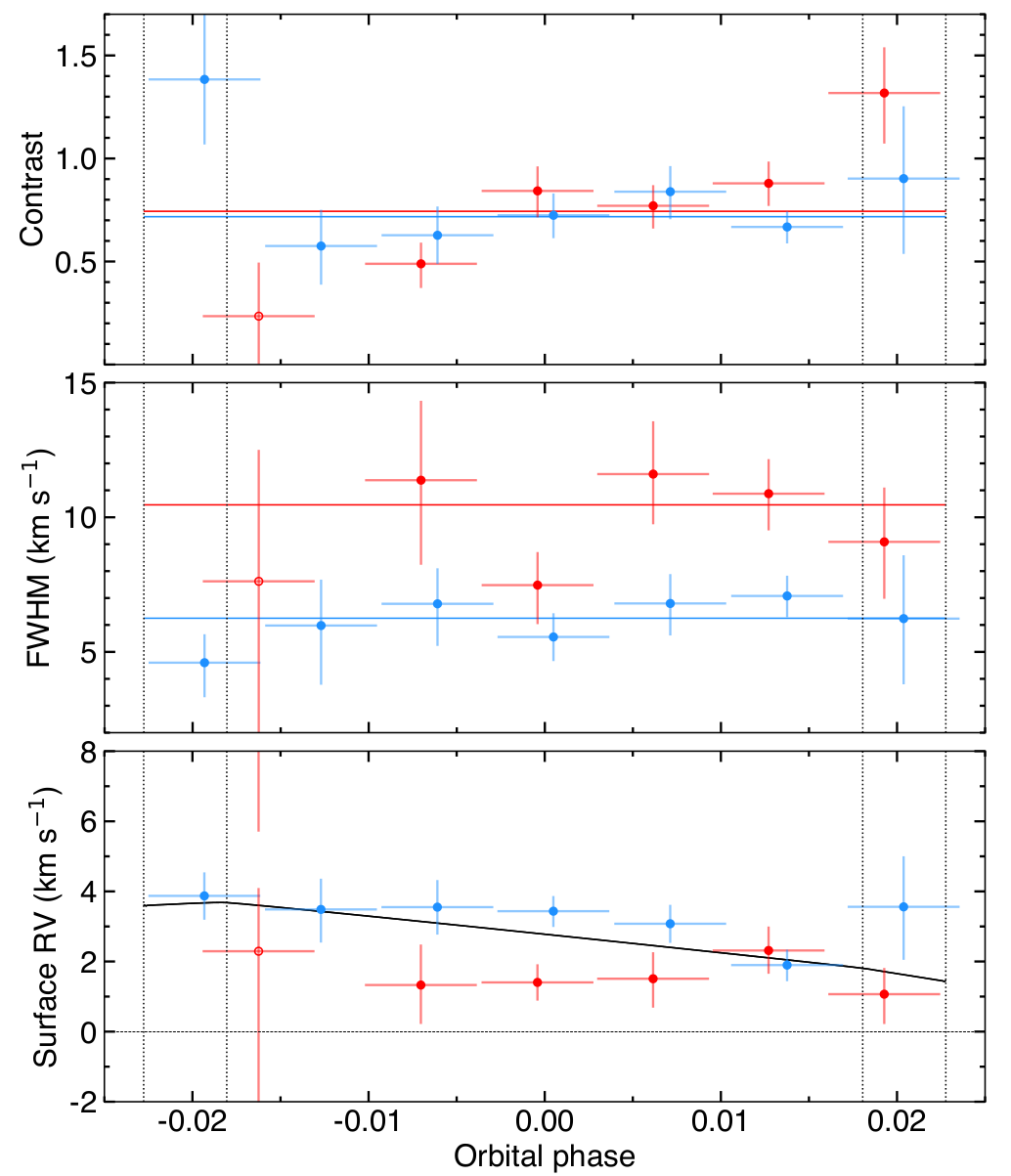}
                \centering
                \caption[]{Properties of the stellar surface regions occulted by \Nplanet, in blue for Visit 1 and
                        red for Visit 2. The dashed vertical lines are the transit contacts. The horizontal bars indicate the
                        duration of each exposure. The vertical bars indicate the 1$\sigma$ HDI intervals. The solid curves are the
                        best models to each property, derived from a joint RMR fitted to both visits (excluding the first
                        exposure in Visit 2). The RV model is common to both visits. The model contrast and FWHM are
                        specific to each visit and are shown here after convolution of the model line by the CORALIE line spread function (LSF).}
                \label{fig:PropLoc_TOI858b}
        \end{figure}

        As shown in Fig.~\ref{fig:CCFintr_map}, the stellar line is clearly detected in most individual CCF$_\mathrm{intr}$
        with narrow and well-defined posterior distribution functions (PDFs) for their model parameters (Figs.~\ref{fig:PDF_V1_indivexp} and
        \ref{fig:PDF_V2_indivexp}), which allowed us to extract the time series of the local stellar line properties 
        (Fig.~\ref{fig:PropLoc_TOI858b}). The only exception is the first exposure in Visit 2, which was excluded from further analysis. 
        Surprisingly, the local RV series displays larger values overall in Visit 1
        than in Visit 2. Nonetheless, both series are positive and constant at the first order, showing that
        \Nplanet\ transits the stellar hemisphere, rotating away from the observer at about the same stellar longitude.
        There is no evidence for centre-to-limb variations in the local stellar line shape, as the local
        contrast and FWHM series remain roughly constant along the transit chord. The width of the local
        line, however, appears to be broader in the second visit, suggesting a possible change in the stellar
        surface properties.

        \subsubsection{Joint transit analysis}
        \label{sec:step3}
        
        In the third step, all CCF$_\mathrm{intr}$ were fitted together with a joint
        model. Based on step two, the local stellar line was modelled as a Gaussian profile
        with a constant contrast and an FWHM along the transit chord but with values specific to each visit. The
        RV centroids of the theoretical lines were set by the surface RV model described in \citet{Cegla2016}
        and \citet{Bourrier2017_WASP8}, assuming solid-body rotation for the stellar photosphere and oversampling each
        exposure to account for the blur induced by the planet motion (a conservative oversampling factor of five was used). The time series of the theoretical
        stellar lines were convolved with the CORALIE instrumental response and then fitted to the
        CCF$_\mathrm{intr}$ in both visits using \textit{emcee} MCMC. The model parameters used as jump parameters for the MCMC were the line contrast
        and FWHM, the sky-projected obliquity $\lambda$, and stellar rotational velocity $v$\,sin\,$i_*$.
        Uniform priors were set on all parameters: over the same range as in step two for the contrast and
        FWHM, between [0 - 30]\,km\,s$^{-1}$ for $v$\,sin\,$i_*$, and over its definition range ([-180 ,
        180]$^{\circ}$) for $\lambda$. 
        
\begin{figure}
                                \includegraphics[trim=0cm 0cm 0cm 0cm,clip=true,width=\columnwidth]{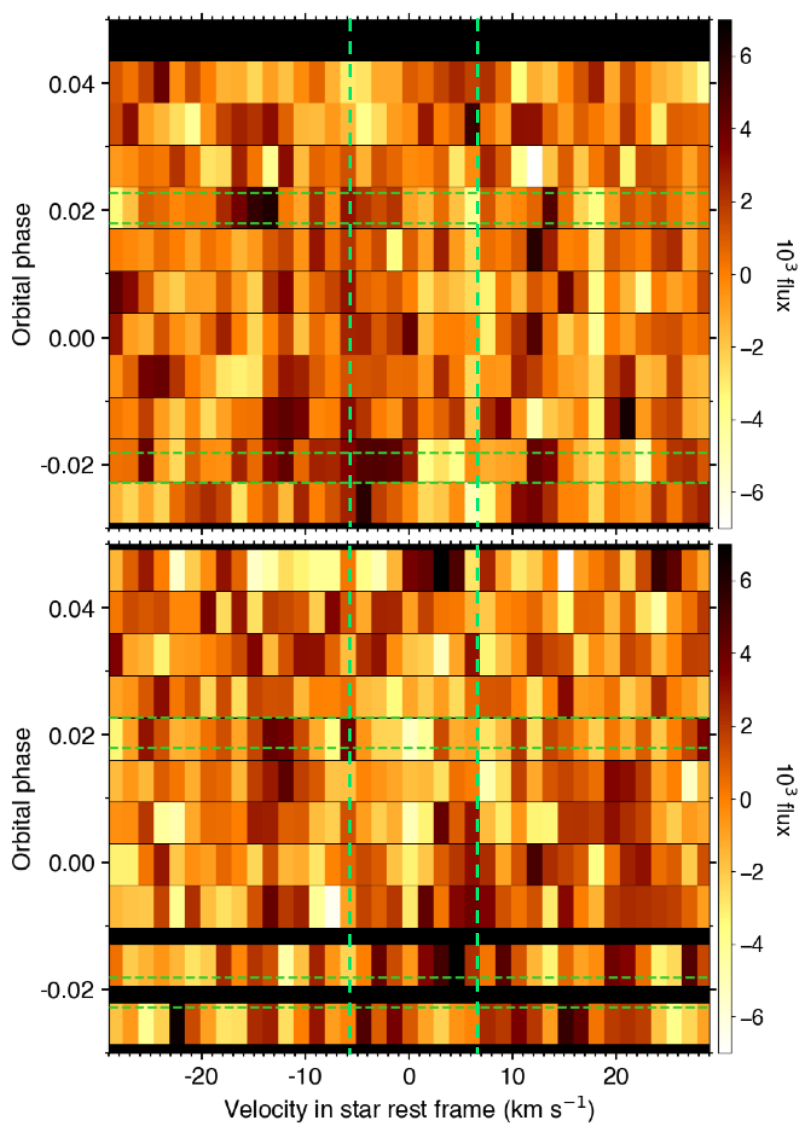}
                \centering
                \caption[]{Maps of the out-of-transit residuals and of the in-transit residuals between
                        CCF$_\mathrm{intr}$ and their best-fit model in Visit 1 (top panel) and Visit 2 (bottom panel).
                        Transit contacts are shown as green dashed lines. The green vertical dashed lines show the spectroscopic, sky-projected stellar rotational velocity. }
                \label{fig:Res_map_TOI858b}
        \end{figure}

        Posterior distribution functions for the model parameters are shown in Fig.~\ref{fig:PDF_TOI858b}. The best fit yielded a reduced
        $\chi^2$ of 1.3 ($\chi^2$ = 634 for 474 degrees of freedom) with models that reproduce the
        local stellar lines along the transit chord well, as can be seen in the residual maps shown in
        Fig.~\ref{fig:Res_map_TOI858b}. Properties of the best-fit line model convolved with the CORALIE
        response are shown in Fig.~\ref{fig:PropLoc_TOI858b}. The local line has a similar contrast in the
        two visits (71.2$\stackrel{+4.8}{_{-4.3}}$\% for Visit 1 and 74.1$\pm$5.2\% for Visit 2), but it is
        significantly broader in the second visit (6.26$\stackrel{+0.25}{_{-0.37}}$\,km\,s$^{-1}$ for Visit
        1 and 10.44$\pm$0.75\,km\,s$^{-1}$ for Visit 2). Whether the origin of this variation is stellar or
        not, we note that $\lambda$ and $v$\,sin\,$i_*$ are not correlated with the contrast and FWHM of the
        fitted lines (Fig.~\ref{fig:PDF_TOI858b}). We derived $v$\,sin\,$i_*$ = 7.09$\pm$0.52\,km\,s$^{-1}$,
        which is larger than the value of $5.80\pm 0.25$\,km\,s$^{-1}$ obtained from a spectroscopic analysis of the CORALIE data. The main result from
        our analysis is the derivation of the projected spin-orbit angle, $\lambda$ =
        99.3$\stackrel{+3.8}{_{-3.7}}^{\circ}$, which shows that \Nplanet\ is on a polar orbit. 
        
        We performed two tests to assess the reliability of this conclusion. First, as the derived local line
        contrasts depend on the accuracy of the transit light curve used to scale the CCF$_\mathrm{DI}$
        (Sect.~\ref{sec:step1}), we thus varied the transit depth of \Nplanet\ within its 3$\sigma$
        uncertainties. We found that it changes $\lambda$ by less than 0.5$^{\circ}$. Then, we fitted the
        two visits independently. We found significant differences between the derived $v$\,sin\,$i_*$
        (8.2$\pm$0.6\,km\,s$^{-1}$ in Visit 1 and 3.6$\pm$0.9\,km\,s$^{-1}$ in Visit 2), neither of which is
        similar with the value derived from the CORALIE CCF-width in Sect. \ref{sec:rot} (5.80$\pm$0.25\,km\,s$^{-1}$). Interestingly, 
         the latter is consistent with the weighted mean of Visit 1 and 2 values (6.8$\pm$0.5\,km\,s$^{-1}$). In the present case 
        of a polar orbit, the rotational velocity is directly constrained by the overall level of the surface RV series, which explains 
        why we derived different values for the two visits (Fig.~\ref{fig:PropLoc_TOI858b}). The physical origin of this difference, 
        however, is unclear. No detrimental contamination was identified in Fibre B of CORALIE,    which was pointing on-sky in Visit 1. It is
        possible that an instrumental drift may have biased the RVs during the visit, which is why we put
        Fibre B on the FP simultaneous reference in Visit 2. The first half of the
        transit in Visit 2 was obtained with lower S/N (down to 12 in the first exposures compared to 17
        afterwards, as measured in order 46), possibly due to a change in sky conditions. This makes it
        difficult to determine which data set is the most accurate, considering that they mainly differ
        during the first half of the transit. The projected spin-orbit angle, however, is less affected by
        these variations than $v$\,sin\,$i_*$ and remains consistent within 1$\sigma$ between the two visits
        (99.4$\stackrel{+3.1}{_{-3.0}}^{\circ}$ in Visit 1 and 80.9$\stackrel{+17.0}{_{-13.7}}^{\circ}$ in
        Visit 2), confirming the misalignment of the \Nplanet\ orbital plane.

 
        To calculate the actual obliquity (as opposed to its sky projection),
        we needed information about the inclination of the stellar rotation axis
        relative to the line of sight. Such information can be obtained from
        the combination of the stellar radius, rotation period,
        and projected rotation velocity ($v\sin i_\star$). In this case,
        the rotation period of the planet-hosting star, TOI-858\,B, is uncertain.
        A photometric period of 6.4 days
        was detected in the TESS and WASP-South light curves, but the period
        might belong to TOI-858\,A instead.
        Bearing this caveat in mind and assuming the period belongs to B,
        we derived constraints on the stellar inclination angle using an MCMC
        procedure, following
         \cite{MasudaWinn2020}. The free parameters were
        $\cos i_\star$, which was subject to a uniform prior;
        $R_\star/R_\odot$, which was subject to a Gaussian
        prior with a mean of 1.308 and standard deviation of 0.038 (see Table~\ref{tab:tableTOI858}); and $P_{\rm rot}$, which was subject
        to a Gaussian prior with a mean of 6.42 days and a standard deviation of 0.64
        days (enlarged to 10\% to account for systematic effects such as
        differential rotation).  The log-likelihood was taken to be
        \begin{equation}
            -\frac{1}{2} \left(
            \frac{
              \frac{2\pi R_\star}{P_{\rm rot}} \sqrt{1-\cos^2 i_\star} -
              5.80\,{\rm km/s}
              }
              {0.25\,{\rm km/s}} 
            \right)^2
        \end{equation}
    based on the CORALIE-based measurement of $v\sin i_\star$.
    The result for $\cos i_\star$ was $0.82^{+0.04}_{-0.05}$.
    We combined this result with the measurements of $\lambda=99.3\pm 3.8$~degrees
    and $i_{\rm o} = 86.8\pm 0.5$~degrees to arrive at two possibilities
    for the stellar
    obliquity: $\psi = 92.7\pm 2.5$ degrees or $98.0\pm 2.5$ degrees (the discrete degeneracy arises because we do not know the
    relative signs of $\cos i_\star$ and $\cos i_{\rm o}$).
    Thus, under these assumptions, the stellar spin axis and normal to the orbital plane are nearly perpendicular.
    
    This conclusion does not depend strongly on the exact constraints
    on $v\sin i_\star$. For example, if we use the constraint
    $v\sin i_\star = 7.09\pm 0.52$~km/s, based on the RM analysis,
    the results for $\psi$ are modified to
    $94.5\pm 3.0$ and $98.9\pm 3.0$ degrees.
    In fact, because $\lambda$ is so well constrained,
    the conclusion that the axes are nearly perpendicular
    does not even depend strongly on the assumption that
    the rotation period is 6.4~days.  When we repeated the calculation
    for any choice of rotation period between 0.1 and 11 days,
    the best-fit value of $\psi$ varied between 87 and 100 degrees.

        \begin{figure}
                \includegraphics[trim=0cm 3cm 0cm 0cm,clip=true,width=\columnwidth]{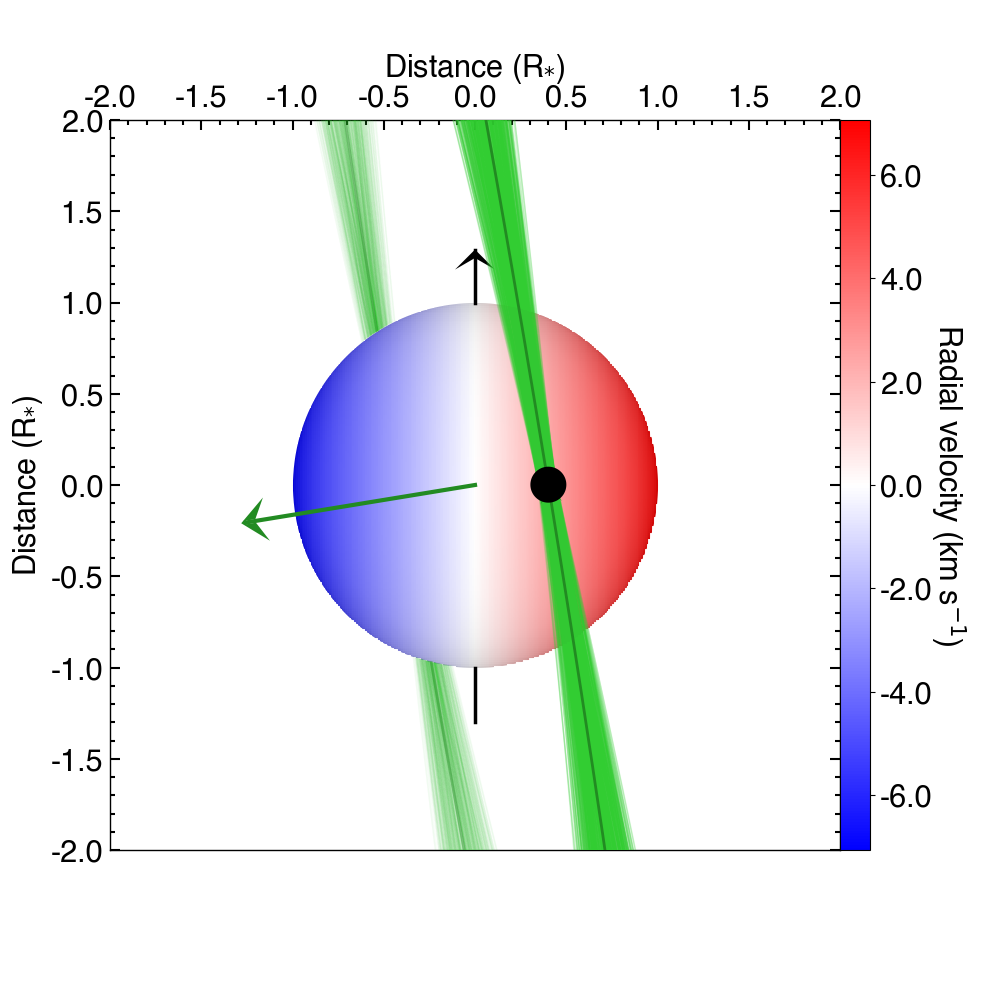}
                \centering
                \caption[]{Projection of \Nstar\ in the plane of sky for the best-fit orbital architecture. The
                        black arrow shows the sky-projected stellar spin. The stellar disc is coloured as a function of its
                        surface RV field. The normal to the orbital plane of \Nplanet\ is shown as a green arrow. The thick
                        green solid curve represents the best-fit orbital trajectory. The thin lines surrounding it show
                        orbits obtained for orbital inclination, semi-major axis, and sky-projected obliquity values drawn
                        randomly within 1$\sigma$ from their probability distributions. The star, planet (black disc), and
                        orbits are to scale.}
                \label{fig:System_view}
        \end{figure}

\subsection{Dynamical analysis}\label{sec:dynamic}

\subsubsection{The \Nstar\ -- \NstarB \;system}
The two stars, \Nstar\ and \NstarB, are separated by $\rho\sim$11\arcsec\ ($\sim$3000 au; \citealt{gaiaEDR3}) and have
similar proper motion and parallax values in Gaia EDR3, suggesting they represent a wide binary pair and
are a good target for examination of angular momentum vector alignment between binary and transiting planet
orbits. However, the escape velocity for a system of two stars with these given masses and uncertainties at a
separation of $\sim$3000~AU is $0.09\pm0.06$~km~s$^{-1}$, while the relative velocity vector given by
EDR3 proper motions and RV for both stars is $3.8\pm0.2$~km~s$^{-1}$, nearly 18-$\sigma$
higher than the escape velocity.  Thus, only unbound, hyperbolic trajectories are consistent with these
relative velocities. 

Both have high quality EDR3 astrometric solutions as measured by the re-normalised unit weight error
(RUWE): \Nstar\ RUWE = 1.009 and \NstarB\ RUWE = 1.0166, where RUWE$\approx$1 is a well-behaved solution
\citep{Lindegren2018RUWE}.\footnote{\url{https://www.cosmos.esa.int/web/gaia/dr2-known-issues\#AstrometryConsiderations}}
The TOI-858~A--B pair is not resolved in Hipparcos or Tycho. There are two additional astrometric
measurements of the pair, both from 1894, in the Washington Double Star Catalogue 
(WDS; \citealt{Mason2001WDS}), with $\rho=11.5\arcsec$ and position angle (PA) $=164^\circ$ in 1894, compared
to $\rho=10.94903\pm1\times10^{-5}\ \arcsec$, 
PA $=169.90649\pm6\times10^{-5}$ deg in Gaia EDR3.
The plane-of-sky relative velocity given by those measurements ($\sim$12 km s$^{-1}$) is larger 
than the Gaia EDR3 relative velocity.  Neither star is in the \cite{El-BadryGaiaBinaries} catalogue of
binaries identified in Gaia EDR3.  Following the method described in \cite{Pearce2021Boyajian} Sect.
3.1.1, we determined the probability of a chance alignment to be $\sim1\times10^{-6},$ given the density
of all objects in Gaia EDR3 within a 10$^\circ$ radius and 1$-\sigma$ of the \Nstar\ proper motion and
parallax. We conclude that \Nstar\ and \NstarB\  either (1) are a formerly bound binary that recently
became unbound, (2) have inaccurate solutions in Gaia EDR3 despite the low RUWE values, or (3) are a chance
alignment, despite the low probability. We can thus conclude that the system is either a binary that recently became unbound (conclusion 1) or a binary that has
inaccurate Gaia EDR3 solutions (conclusion 2).

\subsubsection{Assessment of a Kozai--Lidov evolution}

\begin{figure}
    \centering
    \includegraphics[width=\hsize]{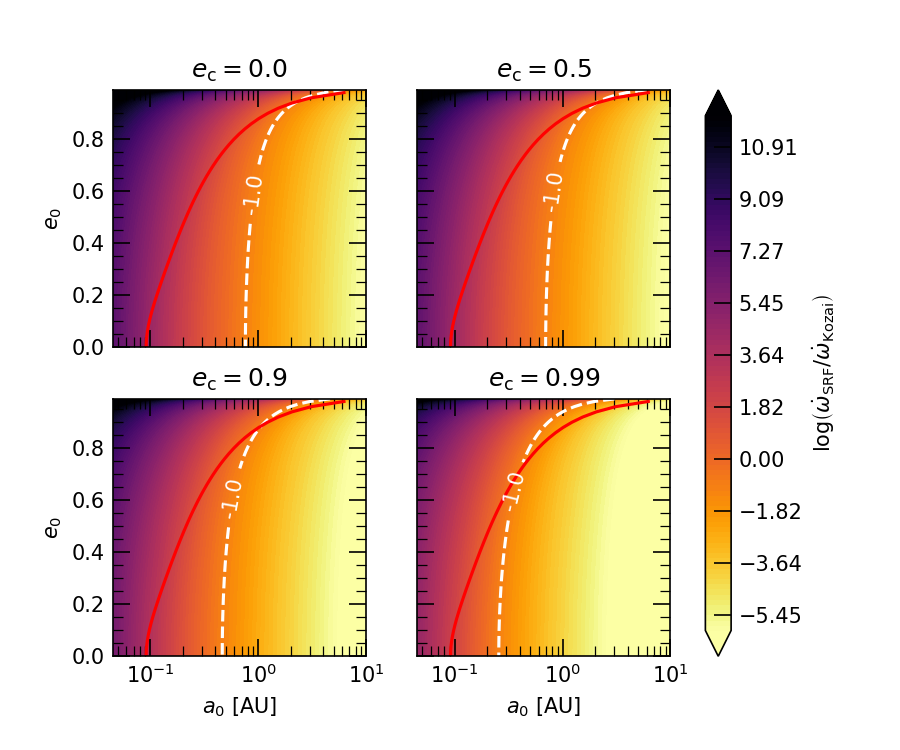}
    \caption{Ratio of short-range forces to Kozai--Lidov precession rate as a function of the initial semi-major axis and 
    eccentricity of \Nplanet{}'s orbit for four different values of \NstarB{}'s orbital eccentricity. The white dashed lines show 
    $\dot{\omega}_{\rm SRF}/\dot{\omega}_{\rm Kozai}=0.1$. The red lines indicate the co-rotation radius.}
    \label{fig:kozai}%
\end{figure}

Given the polar orbit of \Nplanet{}, we examined the possibility of this architecture
being caused by the action of the stellar binary companion. Under some circumstances, a
distant third body can trigger the Kozai--Lidov effect \citep{Kozai62,Lidov62,Naoz16}, a
secular dynamical mechanism that makes the inner orbit's eccentricity and inclination
oscillate. The Kozai--Lidov resonance has been invoked as a possible explanation for
misaligned orbits \citep[e.g.,][]{Fabrycky2007,Anderson16,Bourrier_2018_Nat}, but in some cases it
can be quenched due to short-range forces of the star \citep{Liu15}.

Indeed, some short-range forces induce precession of the periapsis in the direction opposite
of the Kozai--Lidov effect, possibly suppressing the resonance if they are
strong enough \citep[e.g.,][]{Wei21}. Thus, one necessary condition for the onset of the
Kozai--Lidov mechanism is that its associated precession rate $\dot{\omega}_{\rm Kozai}$
must be higher than the precession rate $\dot{\omega}_{\rm SRF}$ of the short-range-forces.
Using the parameters in Tables \ref{tab:starB} and \ref{tab:starA}, we computed the
$\dot{\omega}_{\rm SRF}/\dot{\omega}_{\rm Kozai}$ ratio for a broad range of initial
semi-major axes $a_0$ and eccentricities $e_0$ as well as four different values of the
companion's eccentricity $e_{\rm c}$ so as to investigate in which region of the
parameter space a Kozai--Lidov resonance could be triggered. The short-range
forces we included in $\dot{\omega}_{\rm SRF}$ are general relativity, static tides, and
rotational forces (with the same formalism as in e.g.,~\citet{Eggleton01}).

Figure \ref{fig:kozai} illustrates the results. We show the 
$\dot{\omega}_{\rm SRF}/\dot{\omega}_{\rm Kozai}=0.1$ threshold as a typical value below
which the Kozai--Lidov effect can take place as well as the co-rotation radius beyond
(resp.~inside) which tides widen (resp.~shrink) the orbit of the planet. The parameter
space regions where the Kozai--Lidov mechanism can be strongly active (i.e.,~regions
where the $\dot{\omega}_{\rm SRF}/\dot{\omega}_{\rm Kozai}$ ratio is low) are nearly all
located beyond the co-rotation radius. Hence, even though \Nplanet{} formed with a
favourable orbital configuration for the launch of the Kozai--Lidov effect, reaching the
present-day close-in orbit would have been prevented by tidal forces. Indeed, they would
always increase the semi-major axis, except for extremely high eccentricities of the
binary companion ($e_{\rm c}>0.99$) when favourable architectures can be found inside
the co-rotation radius. We checked the validity of these analytical results with
comprehensive numerical simulations using the \texttt{JADE} code \citep{Attia21} for a
representative subset of the parameter space, and they corroborated our conclusions.

In summary, if the binary companion's eccentricity is not implausibly high, the
Kozai--Lidov scenario can be excluded as a possible explanation for the polar orbit of
\Nplanet{}. As \NstarB{} is far away, compatible Kozai--Lidov effects would be quenched
by short-range forces generated by \Nstar{}.

\section{Discussion and conclusions}\label{sec:discuss}

The planet \Nplanet{} may represent another case of a ``perpendicular planet,'' adding statistical weight to the trend identified by 
\citet{Albrecht21}.
Even though we lack an unambiguous measurement of the stellar inclination in order to disentangle the sky-projected and the true spin-orbit angle, we can still assert the polar nature of the orbit, as $\lambda$ is very well constrained (Sect. 4.1.4). Moreover, the sky-projected and the 3D spin-orbit angle tend to be close when the former is near $90^\circ$ \citep{2009ApJ...696.1230F}.
In any case, because of \NstarB{}'s wide separation, a Kozai--Lidov mechanism is unlikely to be the origin of \Nplanet{}'s highly misaligned orbit. With the current orbital parameters, such an effect raised by the binary companion would be cancelled out by short-range forces between \Nstar{} and its orbiting planet. Other explanations for a polar orbit include stellar flybys \citep[e.g.,][]{Rodet2021}, secular resonance crossings \citep{Petrovich2020}, and magnetic warping \citep[e.g.,][]{Romanova2021}. The flyby scenario would be compatible with the orbit of the companion star \NstarB; however, more precise astrometry would be needed to determine the exact nature of its orbit.
Alternatively, the present wide binary stars TOI-858 A and B could have been closer together in the past, allowing for the high-eccentricity tidal migration to take place. This could explain the origin of the discovered hot Jupiter TOI-858 B b as well as the polar spin-angle misalignment that we measured in this study \citep[]{Vick23}. Theoretical work is needed to assess the feasibility of such a scenario. 
Future observations will reveal if hot  and misaligned giant planets correlate with the presence of a distant stellar companion.

In this publication, we have reported the discovery of a Jovian planet 
transiting \Nstar\ on a polar orbit. The combined analysis of photometric, high-resolution spectroscopic,
 astrometric, and imaging observations has led to the following main results:

\begin{itemize}
        \item From the joint transit photometry and RV analysis of planet \Nplanet\,, we find that it is on a
        $3.2797178 \pm 0.0000014$ day orbit around its 1.08 \msun\ G0 host star. 
        The planet has a mass of \mplanet\ and a radius of $1.255 \pm 0.039$ 
        \rj.
        \item The Rossiter-McLaughlin Revolutions analysis leads to the conclusion that the planet is on a polar orbit 
        with a sky-projected obliquity of $\lambda$ = 99.3$\stackrel{+3.8}{_{-3.7}}^{\circ}$.
        \item Assuming that the photometric periodicity is from the host star, we find that the stellar spin axis and normal to the 
        orbital plane
   are nearly perpendicular, and due to the fact that $\lambda$ is so well constrained,
        this result does not strongly depend on the assumed rotation period.
        \item From our combined RV-astrometry analysis, we conclude that \Nstar\ and \NstarB\ are indeed the two components  of a binary system, and if the Gaia EDR3 solutions are accurate, they recently became unbound.
        \item From our dynamical study, we conclude that Kozai--Lidov can be excluded as a possible explanation for the polar orbit of 
        \Nplanet{}. However, a stellar flyby would be compatible with the current astrometry of the companion star \NstarB.
        
\end{itemize}

\begin{acknowledgements}
We would like to thank the anonymous referee for the very 
constructive comments which significantly improved the scientific quality of the article.
                J.H. is supported by the Swiss National Science Foundation (SNSF) 
                through the Ambizione grant \#PZ00P2\_180098.
                L.D.N thanks the SNSF for support under Early Postdoc.Mobility grant \#P2GEP2\_200044. 
                V.B. is supported by the National Centre for Competence in Research
                “PlanetS” from the SNSF. V.B. and O.A. are funded
                by the ERC under the European Union's Horizon 2020 research
                and innovation programme (project {\sc Spice Dune}, grant agreement No 947634).   
                A.B.D. was supported by the National Science Foundation Graduate Research Fellowship Program under Grant Number DGE-1122492.
        J.V. acknowledges support from the Swiss National Science Foundation (SNSF) under the Ambizione grant \#PZ00P2\_208945.
                Funding for the TESS mission is provided by NASA's Science Mission Directorate. We acknowledge the use of public TESS data from pipelines at the TESS Science Office and at the TESS Science Processing Operations Center. This research has made use of the Exoplanet Follow-up Observation Program website, which is operated by the California Institute of Technology, under contract with the National Aeronautics and Space Administration under the Exoplanet Exploration Program.
 This work has made use of data from the European Space Agency (ESA) mission
 {\it Gaia} (\url{https://www.cosmos.esa.int/gaia}), processed by the {\it Gaia}
 Data Processing and Analysis Consortium (DPAC,
 \url{https://www.cosmos.esa.int/web/gaia/dpac/consortium}). Funding for the DPAC
 has been provided by national institutions, in particular the institutions
 participating in the {\it Gaia} Multilateral Agreement.
 This research has made use of the Washington Double Star Catalog maintained at the U.S. Naval Observatory, 
 of the SIMBAD database, operated at CDS, Strasbourg, France, and of NASA’s Astrophysics Data System 
 Bibliographic Services.

\end{acknowledgements}

\bibliographystyle{aa}
\bibliography{bibs/main_bib}

\begin{appendix}
\section{Radial velocity measurements of \Nstar\ and \NstarB}

\begin{table}
        \centering
        \caption{\label{tab:rv_coralie_toi858} CORALIE radial velocity follow-up observations of \Nstar. }       
        \begin{tabular}{cccc}
                \hline\hline
                \noalign{\smallskip}
                MBJD   &   RV   &   $\sigma$ RV   &   BIS \\
                &   [km/s]   &  [m/s]    &   [m/s]    \\
                \noalign{\smallskip}
                \hline
                \noalign{\smallskip}
                58708.871832  &  64231.37  &  20.17  &  -37.1 \\
                58739.911963  &  64496.23  &  19.69  &  -35.46\\
                58742.862604  &  64491.79  &  19.88  &  -16.92\\
                58753.894008  &  64334.29  &  35.2  &   -53.5\\
                58754.85256  &  64249.83  &  31.49  &   -3.91\\
                58755.82014  &  64465.68  &  26.27  &   12.95\\
                58756.84846  &  64434.94  &  28.40  &   12.59\\
                58757.702855  &  64276.99  &  32.95  &  -14.46\\
                58758.755781  &  64358.17  &  34.99  &  -49.44\\
                58803.771321  &  64198.76  &  19.28  &  -37.28\\
                58816.714463  &  64237.19  &  15.87  &  -42.44\\
                58820.530409  &  64233.82  &  41.35  &  -35.93\\
                58822.570083  &  64321.24  &  102.41  & -57.23\\
                58822.743655  &  64317.07  &  104.37  & -30.22\\
                58822.765321  &  64365.66  &  104.35  & -9.96\\
                58822.786895  &  64353.17  &  103.9  &  -1.53\\
                58831.666624  &  64493.16  &  22.86  &  26.41\\
                59232.536537  &  64417.76  &  33.48  &  48.12\\
                59232.705097  &  64366.5  &  25.69  &   3.62\\
                59232.726763  &  64324.67  &  27.09  &   62.2\\
                59232.748591  &  64310.16  &  29.27  &  -24.19\\
                59232.770245  &  64371.40  &  35.14  &  -74.25\\
                \noalign{\smallskip}
                \hline
                \noalign{\smallskip}    
        \end{tabular}
\end{table}

\begin{table}
        \centering
        \caption{\label{tab:rv_chiron_toi858} CHIRON radial velocity follow-up observations of \Nstar. }       
        \begin{tabular}{ccccc}
                \hline\hline
                \noalign{\smallskip}
                MBJD  &  RV  &  $\sigma$ RV  &  FWHM  &  BIS\\
                &   [m/s]   &  [m/s]  &  [m/s]  &  [m/s]  \\
                \noalign{\smallskip}
                \hline
                \noalign{\smallskip}
                58708.9269  &  -63  &  41.5  &  12.239  &  -40.7\\
                58709.92875  &  250.2  &  31.9  &  12.446  &  31.8\\
                58710.91343  &  230.3  &  41.9  &  11.572  &  311.5\\
                58713.9308  &  205.9  &  17.6  &  12.637  &  -60.2\\
                58718.91475  &  -43.2  &  34.8  &  12.153  &  -50.1\\
                58727.89385  &  -71.7  &  23.8  &  12.918  &  -54.3\\
                58728.89541  &  0  &  26.8  &  12.742  &  271.8\\
                \noalign{\smallskip}
                \hline
                \noalign{\smallskip}    
        \end{tabular}
\end{table}

\begin{table}
        \centering
        \caption{\label{tab:rv_coralie_tyc} CORALIE radial velocity follow-up observations of \NstarB. }      
        \begin{tabular}{cccc}
                \hline\hline
                \noalign{\smallskip}
                MBJD  &  RV  &  $\sigma$ RV &  BIS \\
                &  [m/s] &  [m/s]  &  [m/s] \\
                \noalign{\smallskip}
                \hline
                \noalign{\smallskip}
                58816.744983  &  65519.52  &  13.74  &  -7.19\\
                58822.809915  &  65525.80  &  26.27  &  702.8 \\
                58831.681392  &  65505.75  &  23.41  &  117.9 \\
                58843.597684  &  6551.71  &  19.63  &   72.6 \\
                58860.658389  &  65502.95  &  23.92  &  -283.6\\
                58899.553766  &  65517.16  &  16.77  &  91.4\\
                58920.523756  &  65554.78  &  25.95  &  879.4\\
                58773.64871  &  65552.58  &  36.95  &   21.2 \\
                58777.737748  &  65572.18  &  33.26  &  429.6 \\
                \noalign{\smallskip}
                \hline
                \noalign{\smallskip}    
        \end{tabular}
\end{table}

\FloatBarrier

\section{Correlation diagrams for the global RMR analysis}

                \begin{figure*}
                        \begin{minipage}[h!]{0.8\textwidth}
                                        \includegraphics[trim=0cm 0cm 0cm 
                                0cm,clip=true,width=0.95\columnwidth]{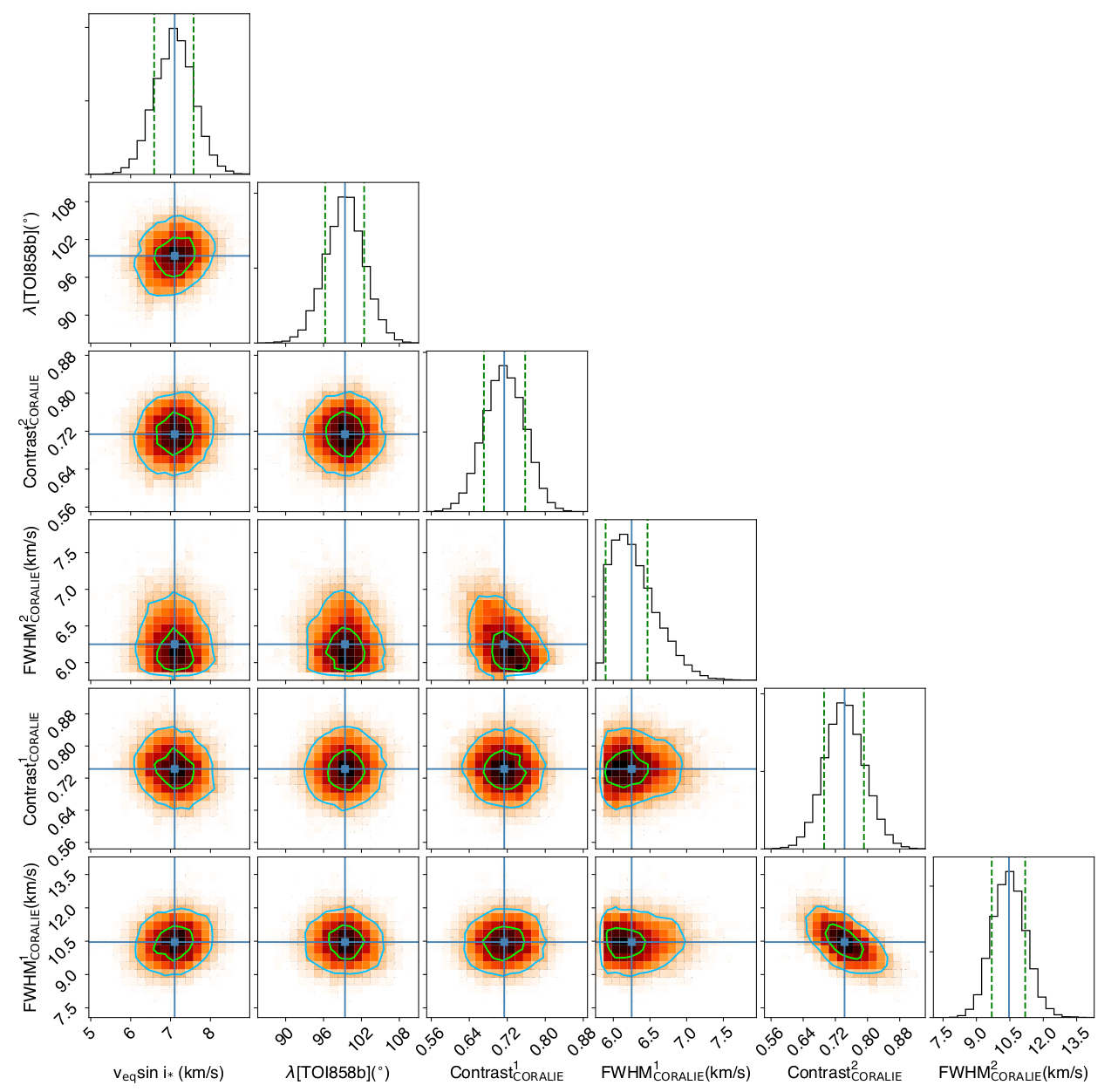}
                                \centering
                        \end{minipage}
                        \caption[]{Correlation diagrams for the PDFs of the RMR model parameters. The CORALIE contrast and
                                FWHM have been derived from the corresponding jump parameters to allow comparison with the observed
                                line. The green and blue lines show the 1 and 2$\sigma$ simultaneous 2D confidence regions that contain,                               respectively, 39.3\% and 86.5\% of the accepted steps. The 1D histograms correspond to the distributions
                                projected on the space of each line parameter, with the green dashed lines limiting the 68.3\% HDIs.
                                The blue lines and squares show median values.}
                        \label{fig:PDF_TOI858b}
                \end{figure*}

\begin{figure*}
                \begin{minipage}[h!]{\textwidth}
                        \includegraphics[trim=0cm 0cm 0cm 
                        0cm,clip=true,width=0.96\columnwidth]{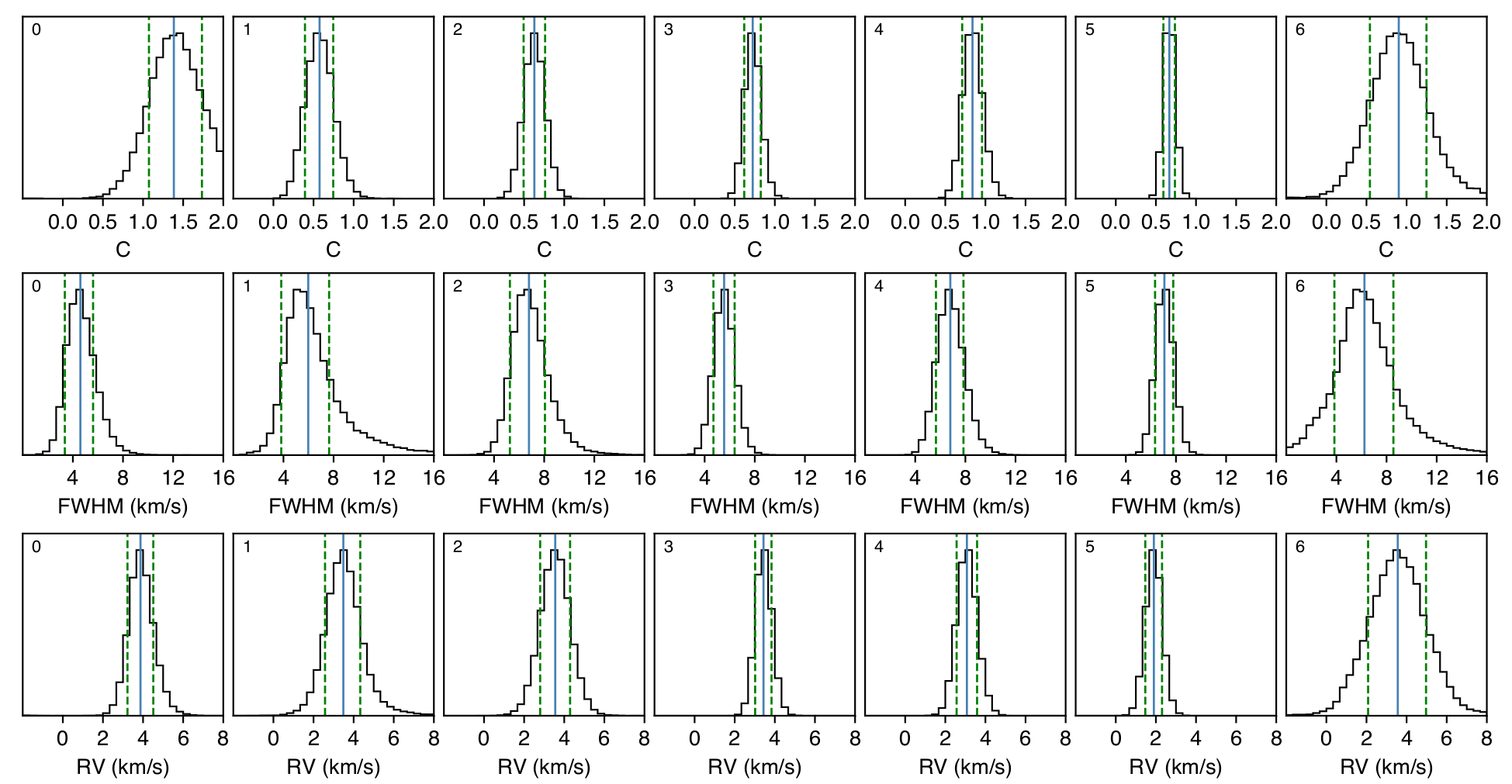}
                        \centering
                \end{minipage}
                \caption[]{Posterior distribution functions of the contrast (upper panels), FWHM (middle panels), and RV centroids (lower
                        panels) of the Gaussian line model fitted to individual CCF$_\mathrm{intr}$ in Visit 1. The deep blue
                        lines indicate the PDF median values, and the green dashed lines show the 1$\sigma$ highest density
                        intervals. In-transit exposure indexes are shown in each subplot.}
                \label{fig:PDF_V1_indivexp}
        \end{figure*}
        
        \begin{figure*}
                \begin{minipage}[h!]{\textwidth}
                        \includegraphics[trim=0cm 0cm 0cm 
                        0cm,clip=true,width=0.96\columnwidth]{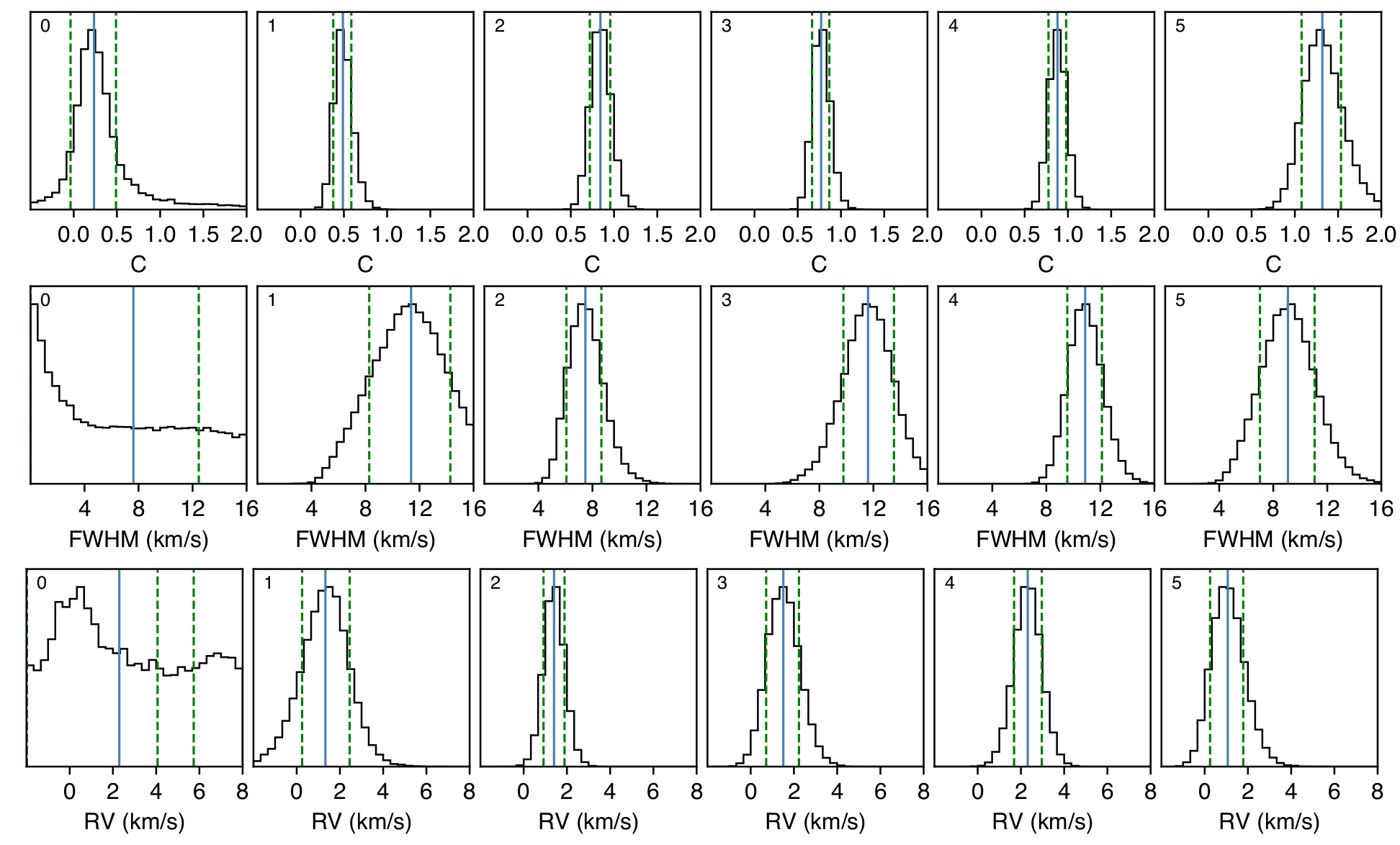}
                        \centering
                \end{minipage}
                \caption[]{Same as Fig.~\ref{fig:PDF_V1_indivexp} but for Visit 2.}
                \label{fig:PDF_V2_indivexp}
        \end{figure*}
        
        \end{appendix}

\end{document}